 \documentclass[preprint,5p]{elsarticle}

\usepackage{amssymb}
\usepackage{amsmath}
\setcitestyle{square}
\usepackage{xcolor}
\usepackage{url}
\usepackage{float}
\usepackage{multirow}
\usepackage{comment}
\usepackage{svg}
\usepackage{listings}
\usepackage{pgfplots}
\usepackage{pgfplotstable}
\usepgfplotslibrary{fillbetween} 
\pgfplotsset{compat=1.18}

\definecolor{backcolour}{rgb}{0.95,0.95,0.92}

\lstdefinestyle{swrl}{
backgroundcolor=\color{backcolour},
breaklines=true,
basicstyle=\ttfamily\small,
captionpos=b,
}

\newtheorem{definition}{Definition}

\journal{Future Generation Computer Systems}

\begin{document}

\begin{frontmatter}



\title{A Context-Aware Knowledge Graph Platform for Stream Processing in Industrial IoT}

\author{Monica Marconi Sciarroni}
 \ead{monica.marconi@univpm.it}
\author{Emanuele Storti\corref{cor1}}
 \ead{e.storti@univpm.it}
 
 \cortext[cor1]{Corresponding author}
 \affiliation{organization={Department of Information Engineering, Polytechnic University of Marche},
            addressline={via Brecce Bianche}, 
            postcode={60131}, 
            city={Ancona},
            country={Italy}}



\begin{abstract}
Industrial IoT ecosystems bring together sensors, machines and smart devices operating collaboratively across industrial environments. These systems generate large volumes of heterogeneous, high‑velocity data streams that require interoperable, secure and contextually aware management. 
Most of the current stream management architectures, however, still rely on syntactic integration mechanisms, which result in limited flexibility, maintainability and interpretability in complex Industry 5.0 scenarios.
This work proposes a context‑aware semantic platform for data stream management that unifies heterogeneous IoT/IoE data sources through a Knowledge Graph enabling formal representation of devices, streams, agents, transformation pipelines, roles and rights. 
The model supports flexible data gathering, composable stream processing pipelines, and dynamic role‑based data access based on agents' contexts, relying on  Apache Kafka and Apache Flink for real‑time processing, while SPARQL and SWRL-based reasoning provide context‑dependent stream discovery. 
Experimental evaluations 
demonstrate the effectiveness of combining semantic models, context‑aware reasoning and distributed stream processing to enable interoperable data workflows for Industry 5.0 environments.

\end{abstract}




\end{frontmatter}



\section{Introduction}
\label{sec:introduction}

The ongoing digital transformation of industrial environments is increasingly driven by the pervasive deployment of cyber-physical systems, among which a relevant role is played by data-driven and intelligent systems enabled by the Industrial Internet of Things (IIoT). This term, along with its evolution as ``Internet of Everything'' (IoE), 
stands for the large-scale integration of sensing, actuation, and communication capabilities in industrial ecosystems, enabling more advanced and smarter automation, monitoring and optimization.
At the same time, such a massive number of interconnected devices (among which traditional sensors, wearable devices, smart objects) continuously generate data at different rates, rely on diverse communication protocols (e.g., MQTT, CoAP, HTTP) and expose data schemas that are typically non-uniform. This heterogeneity introduces significant complexity in data management, analysis and governance, making it difficult to realize integrated and robust industrial applications.
These challenges become even more critical in the context of Industry 5.0 \cite{leng2022industry}, which replaces static configurations and predefined roles with human-centric, sustainable, and resilient processes where human agents and smart devices are able to seamlessly interact and cooperate in a flexible, secure way, adapting to changing contexts and conditions.
These challenges emerge in practical industrial operations, as highlighted by the following real-world scenarios:
\begin{itemize}
\item 
A smart HVAC (Heating, Ventilation, and Air Conditioning) system monitors $CO_2$ levels in Area A2, in order to increase ventilation if the concentration exceeds 1000 ppm. To do so, it must identify all sensors capable of measuring $CO_2$, despite the fact that they use different communication protocols (e.g., MQTT vs HTTP), expose heterogeneous schemas, and apply incompatible naming conventions (e.g., $CO2$ vs $Carbon\_dioxide$ vs $CO2\_level$.

\item 
Marco is a Junior Operations Engineer responsible for maintenance of all sensors and systems deployed within the industrial Facility A. 
To quickly respond to a critical situation,  
he temporarily asks the collaboration of Anne, a Senior Operation Engineer, for the emergency restoration of the production line A2/1. Hence, she must immediately gain access to the real-time streams produced by those sensors. However, traditional access-control mechanisms tied to static device identifiers cannot easily support such short-term, context-de\-pen\-dent delegation.

\item 
In an assembly station, a collaborative robotic arm works with a human operator to assemble precision components. The robot is equipped with a torque sensor (500 Hz) to monitor joint stress, and an infrared thermal sensor (1 Hz) to detect overheating in actuators. A pipeline processes these heterogeneous streams to detect local anomalies, e.g., when the average torque per second exceeds a predefined warning threshold, then merges the streams and performs a context-aware aggregation of the anomalies considering the task at hand, to trigger adaptive robot behavior and alert the operator if needed.
As different employees can access the machine under different contextual conditions, configuring access policies and updating the pipeline over time becomes increasingly difficult.
\end{itemize}

These cases illustrates the importance of solutions 
for syntactic and semantic interoperability that can support automatic discovery and uniform interpretation of relevant streams, helping agents (both employees and autonomous smart objects or co-bots) in retrieving relevant information under role-based access policies and dynamic relations of delegation/collaboration.
They also motivate the need for well 
documented representation of streams and transformation pipelines that supports inspection, auditing, access and long-term maintainability.
In this sense, declarative kno\-wledge representation solutions have proven useful to provide unified and machine-interpretable representations of heterogeneous  resources, which in the industrial context include devices, agents, their roles, rights and relations, enabling shared understanding and consistent interpretation across systems and stakeholders.

The following research questions stem form the above-mentioned challenges: 
\emph{RQ1) To what extent can a declarative, integrated representation of agents, data streams and processing logic support semantic interoperability across heterogeneous IIoT devices?}, \emph{RQ2) How can  semantic reasoning be leveraged to dynamically determine context and role-based access rights for data streams?},  
\emph{RQ3) To what extent do semantic context-aware approaches affect the computational overhead and responsiveness of 
Industrial IoT stream processing systems?}.

This work addresses these questions by proposing a context-aware  approach, based on a declarative, formal representation of data streams, processing logic, and access policies in the form of a Knowledge Graph (KG). 
On top of it, a platform for data gathering, processing and storage relies on the Graph to provide semantic interoperability across devices, unify heterogeneous schemas, and enable context- and role-based access control.
Reasoning functionalities take into account the context of an agent, in terms of location, performed activity, role and rights, to dynamically determine accessible data streams and support platform functionalities.
%
%
%
The novel contributions of this work are multi-fold:
\begin{itemize}
    \item A comprehensive Knowledge Graph model is proposed to formally represent relevant Industrial IoT concepts.
    The Graph integrates and extends a set of ontologies (SOSA/SSN, BOT, ORG) allowing the formalization of devices, streams, processing pipelines, and the agent's context (related to RQ1).
    \item A stream gathering and management platform is developed, leveraging the Knowledge Graph to dynamically support stream gathering, processing, di\-stri\-bu\-ted storage, context-based and role-based access for monitoring and federated querying. The platform implementation is based on micro-services and employs Apache Kafka and Flink for low-latency stream handling, while SPARQL queries and SWRL-based reasoning are performed for information extraction and  inference (related to RQ2).
    \item A thorough experimentation is conducted to assess the efficiency of the approach in terms of execution time, latency and scalability, and to evaluate the impact of the semantic layers (KG and reasoning) on the operations. The results demonstrate the feasibility and effectiveness of the proposed framework in realistic IIoT scenarios (related to RQ3).
\end{itemize}
A preliminary version of this approach was presented in \cite{sciarroni2024monitoring}. The present work significantly extends that contribution by providing a more comprehensive definition of the underlying data model, including an extended schema for the Knowledge Graph and reasoning capabilities, along with a more extended discussion of the platform and a thorough experimental evaluation.




The rest of this work is organized as follows: Section \ref{sec:related} discusses relevant related work on semantic approaches for modeling, processing and monitoring data in industrial contexts. Section \ref{sec:definitions} introduces the foundational concepts and definitions underlying the proposed approach, while the semantic data model is discussed in Section \ref{sec:metadata}. Section \ref{sec:framework} is devoted to present the  context-aware processing platform, while Section \ref{sec:reasoning} describes the reasoning services. Section \ref{sec:evaluation} reports the evaluation results, and finally, Section \ref{sec:conclusion} summarizes the work and outlines directions for future research.

\section{Related Work}
\label{sec:related}
This section reviews key contributions in four major areas relevant to this objective: (1) semantic models that provide formal representations entities and data streams  in IoT/IoE scenarios, (2) semantic-based approaches to monitor data streams, (3) frameworks and architectures developed to operationalize these models in real-world environments, and (4) solutions for managing and processing data streams in dynamic and distributed settings.

\subsection{Semantic data models for IoT}
The integration of IoT/IoE technologies presents a significant challenge due to their inherent heterogeneity and the need to combine data streams originating from diverse sources, such as sensors, human actors, and business processes.
To ensure seamless communication and interoperability across heterogeneous systems, in recent years, both academia and industry have increasingly adopted Knowledge Graphs-based solutions to build flexible and homogeneously integrated systems \cite{hogan2021knowledge}. They are indeed recognized for their capability to enhance data integration and  management \cite{li2021exploiting}, supporting semantic reasoning and real-time contextual analysis in Industrial IoT (e.g., in \cite{liu2022knowledge} production and business processing data are considered to enable intelligent decision-making, while  a process-aware graph in \cite{diamantini2023process} supports process-aware analysis).

Significant work has been devoted to develop ontologies that serve as schema foundations for Knowledge Graphs, enabling the representation of device characteristics and their interconnections. 
These ontologies must be sufficiently broad to model networks of sensors, their functional properties and capabilities and  applications built upon them. Moreover, they need to account for the rapid evolution of the IoT domain, where new devices are continuously released, and the substantial variations in services and requirements across different industrial domains  \cite{schlenoff2013literature}.

%
The Semantic Sensor Network (SSN) ontology \cite{compton2012ssn} is recognized as one of the most prominent ontologies for IoT and domotics. Several work built lightweight semantic models on top of the SSN ontology, e.g. \cite{bermudez2016iot,JANOWICZ20191}.
In particular, the IoT-Lite ontology 
\cite{bermudez2016iot} builds a core model containing only the main concepts to support the most standard queries for IoT solutions.
Originally proposed by the W3C Semantic Sensor Network Incubator group, the SSN ontology has also been revised by \cite{JANOWICZ20191} in the Sensor, Observation, Sample, and Actuator (SOSA) ontology. 
This proposal aims at a lightweight vocabulary including broader concepts with respect to the SSN ontology, with the idea to provide a core model that can be integrated and aligned to other specifications.
Researchers also focused on modeling the environments in which sensors are deployed, which are important aspects for Industry 5.0. Examples include DogONT  \cite{bonino2008dogont} for home environments, 
the Building Topology Ontology (BOT) \cite{rasmussen2021bot} to support the exchange of information related to building life-cycles, or the Organization Ontology (ORG) \cite{world2014organization} modeling organizational structures and related information through the concepts of organizations, their actors, activities and roles.

While extensive research has addressed IoT and industrial contexts, the emergence of Industry 5.0 and the Internet of Everything (IoE) presents novel challenges that remain largely unexplored, including the interconnection of people, data, and processes in a unified framework.


\subsection{Semantic monitoring ad processing}
Semantic monitoring has evolved from early ontology-based metadata annotation to advanced context-aware techniques.
\cite{elsaleh2019} developed IoT-Stream, a lightweight semantic model for annotating, querying, and analyzing high-frequency data streams in resource-constrained industrial contexts. 
Similarly, \cite{iiot2023} presented a multi-stage architecture that adapts to concept drift in Industrial IoT (IIoT) environments, ensuring robust monitoring in evolving Industry 5.0 scenarios. \cite{akanbi2020} demonstrated how a distributed stream processing middleware framework is efficient to analyze heterogeneous environmental monitoring data in real-time, improving forecasting accuracy for drought prediction. 
\textcolor{black}{
Further work includes a lightweight ontology \cite{marshoodulla2023watermeter} to annotate IoT sensor data, converting internal data structures to interoperable RDF data and introducing real-time semantic enrichment of MQTT protocols.
The ``Semantic Subscription'' (SemSub) \cite{piller2020semsub} achieves syntactic decoupling through ontology-based search, identifying semantic relationships and proposing optimal topic matches. 
To address inaccuracies in MQTT's semantic representation, a structured framework called ``RulE-BasEd WEb Editor for Semantic-aware Topic Naming in MQTT (MQTT-4EST)'' \cite{piller2022mqtt4est} provides a user-friendly GUI and a topic tree service to ensure valid topic structures, guiding users with ontology-based suggestions.}
Semantic technologies have also been applied to enhance machine learning workflows. \cite{zhou2021semml} proposed SemML, a platform that applies semantic annotation in the development of machine learning models for condition monitoring, reducing manual configuration and improving model reusability.
To support adaptive process management in Industry 4.0, \cite{giustozzi2023} developed a semantic monitoring framework based on the COInd4 ontology, aiming to better integrate sensor data with contextual knowledge through stream reasoning. 
These efforts highlight the potential of semantic models to enhance interoperability of IoE systems, although existing solution mostly focus on annotation of data streams with no direct support for stream processing or manipulation.

\subsection{Frameworks for IoT/IoE}
The evolution of the IoT and its conceptual expansion into the IoE marks a fundamental paradigm shift from connecting physical devices to integrating people, processes, data and things into a unified, intelligent ecosystem. This expanded vision has led to the development of architectural frameworks addressing core technical challenges in terms of semantic interoperability, real-time data processing, energy efficiency, security, scalability, and context-awareness, while also adapting to a wide range of application domains (e.g. smart cities, Industry 4.0, agriculture, healthcare, and environmental monitoring) \cite{miraz2015}. 
Early foundational work established layered architectural models to enable communication, data sharing, and basic service integration across heterogeneous environments. \cite{atzori2010} proposed a conceptual three-layer model, with perception, network, and application layers, separating data acquisition, transmission, and service delivery. Their model emphasized structural integration between the physical and digital domains, and laid the groundwork for interoperability and scalability in IoT systems. Building upon this, \cite{gubbi2013} introduced a cloud-centric architectural framework integrating real-time analytics, modularity, and semantic-aware data fusion. 
As flexible needs grew, \cite{cirillo2019embracing} developed a hyper-connected architecture for cross-domain operation (e.g., smart cities, autonomous driving, healthcare), establishing the conceptual groundwork for the progressive shift toward IoE architectures by promoting interoperability, dynamic scalability, and cross-domain adaptability. 

The concept of IoE has the ambition to integrate diverse IoT specializations, referred to as "IoXs" where "X" denotes a specific domain, into a unified interoperable eco\-system. Building on this idea, \cite{akan2023ioe} proposed a layered IoE architecture in order to facilitate seamless interaction among different IoXs, addressing key challenges of scalability and interoperability, while considering autonomous entities and real-world applications.
Recent research translates the IoE vision into domain-specific applications: \cite{bellini2021ioe} developed a full-stack solution for urban transport resilience in smart cities using big multimedia data; \cite{babar2024sustainable} demonstrated precision farming at biological and molecular levels, thus improving crop monitoring, disease control, and resource optimization; \cite{omalla2024} extended smart home systems integrating actuators, advanced data analytics, and remote access, thereby upgrading automation, energy efficiency, and user experience. 


\subsection{Frameworks for stream processing}
Industry 5.0 has increased the need for up-to-date insights derived from continuous data streams, so real-time stream processing has become a foundational component in IIoT architectures. While the landscape for stream handling is dominated by Apache Kafka, a distributed event streaming platform designed to manage high-volume data with high durability and low latency,
a variety of distributed, various open-source stream processing platforms have emerged. As witnessed by several surveys, these platforms, among which Apache Flink, Storm and Spark streaming, 
provide the computational logic to analyse and process data stream.
A comparative evaluation  is proposed in \cite{nasiri2019evaluation}, based on latency, throughput, and scalability, declaring that Apache Storm and Apache Flink are better suited for applications demanding low-latency performance, while Apache Spark Streaming may be preferable for throughput-intensive scenarios.
In a similar way, \cite{sahal2020big} provided a comprehensive review of open-source stream processing solutions for Industry 4.0 predictive maintenance, emphasizing that a platform selection should consider architectural flexibility and interoperability beyond performance metrics.
Despite the effectiveness in handling high-velocity data, most platforms rely on syntactic data manipulation and lack deep contextual understanding. This has motivated recent research efforts towards enriching stream processing pipelines with semantic models and graph-based approaches. 
Among them, \cite{yemson2023ontology} developed an ontology-based framework for real-time air quality monitoring, integrating contextual knowledge about patient activities and environmental data, their system improved the accuracy of detecting pollution incidents events and asthma attacks, while \cite{casillo2024} proposed a multi-level graph-based approach combining ontologies, context trees and probabilistic models (e.g. Bayesian networks) to improve scenario management in dynamic IoT environments. 

Despite these effort, a full integration of semantic stream processing in industrial IoT/IoE environments remains an open research area.

\section{Foundational concepts}
\label{sec:definitions}

In this section, we discuss the main design principles and provide formal definitions of the core concepts underpinning our semantic context-aware approach for stream management.

\subsection{Design principles}
\label{subsec:design_principles}
Our approach is grounded in the need for semantic interoperability, modular composition, and secure data sharing in Industry 5.0 environments. To address the complexity and scale of modern industrial data systems, we adopt a uniform and extensible model for stream sources and transformations.
 The key design principles that underpin our approach are outlined as follows:


\begin{itemize}
    \item \emph{Everything-as-a-stream}: every platform component is conceptualized as a stream generator, whether they are sensors, smart objects or IT systems. This abstraction provides a unified foundation for data integration and decouples producers, transformers, and consumers. As a result, system components can evolve independently, improving modularity, scalability, and maintainability.

\item \emph{Unique stream identification}: every stream is treated as a first-class entity with a globally unique identifier and associated metadata. This enables consistent referencing and reasoning across the system and supports traceability.

\item \emph{Uniform stream operations}: all transformations are expressed as modular stream operators with typed input and output interfaces. This promotes type safety, enables composition of complex workflows from simple building blocks, and simplifies both execution and reasoning over dataflows.

\item \emph{Composable stream pipelines}: stream processing pipe-lines are modeled as directed acyclic graphs (DAGs), where nodes represent transformation operators and edges represent data streams. The declarative, graph-based model supports formal analysis of data dependencies and execution order. A reusable operator repository maintains the definitions, configurations, and contracts for each operator. Users may extend or customize the repository by adding new operators, promoting adaptability to domain-specific requirements and evolving application needs.

\item \emph{Role-based access control (RBAC)}: to ensure data security and controlled access, role-based access control is enforced by fine-grained roles and permissions at the stream level. 
This ensures that agents (employees, smart objects, co-bots or services) can only access streams they are authorized to use, supporting compliance, protecting sensitive data, and maintaining operational integrity  within collaborative environments.
\end{itemize}

These principles serve as the foundation for the formal model defined in the next subsection and its subsequent implementation via Knowledge Graphs in Section \ref{sec:metadata}.

\subsection{Definitions}
\label{subsec:definitions}
 A key construct in the model is the stream source, defined by an identifier, a type (e.g., sensor, wearable), the location where it is deployed (e.g., a production line, a warehouse), the schema of the messages it produces and a set of metadata.
 
\begin{definition}[Stream Source]
Let $I_S$ be a countable set of unique stream source identifiers, $ \Theta$ be the global set of source types, $\Lambda$ be the global set of spatial descriptors, $\mathcal{N}$ be the universe of attribute names. A stream source is a tuple
\[
\text{f} = \langle id_f, \theta_f, \sigma_f, M_f, \lambda_f \rangle
\]
where $id_f\in I_S$, $\theta_f \in \Theta$, 
$\sigma_f = \langle n_1, n_2, \dots, n_k \rangle$ is the schema of the tuples emitted by $f$, defined as an ordered set of attribute names, where $n_i\in \mathcal{N}$, 
 $M_f$ is a finite map of descriptive metadata as pairs key-value, and $\lambda_f \in \Lambda$.
\end{definition}
We denote by $\mathcal{F}$ the set of all stream sources.
The semantics and typing of attributes of $\sigma_f$ are defined by a set of canonical properties $\mathcal{L}$ (i.e., global property names), a function $\mu:\mathcal{N}\rightarrow \mathcal{L}$ that assigns each local attribute names
to a canonical property and a typing function $\nu:\mathcal{L}\rightarrow \mathcal{D}$ that associates each canonical property with a type domain.

For instance, consider two sensors $f,q\in \mathcal{F}$, with \emph{temp}$ \in \sigma_f$ and \emph{t\_celsius}$ \in \sigma_q$. Both attributes can be mapped to the same property: $\mu$(\emph{temp})=\emph{tempera\-tu\-re\_C} and $\mu$(\emph{t\_cel\-sius}) = \emph{temperature\_C}. The typing function $\nu$(\emph{tempe\-ra\-ture\_C})=$\mathbb{R}$ indicates that the temperature values are real numbers.
The associated data domain of $f$ is denoted by $D_f = \langle \nu(\mu(n_1)),  \cdots, \nu(\mu(n_k) \rangle$, e.g. if $\sigma_f=\langle \emph{temp},\emph{humidity}\rangle$, then $D_f= \langle \mathbb{R}, \mathbb{R}^+\rangle$.

\begin{definition}[Stream]
Let $T$ be a totally ordered set representing the time domain. Given a stream source $f \in \mathcal{F}$, a stream $S_f$ generated by $f$ is a (potentially infinite) function
\[
S_f : \mathbb{N} \to T \times D_f, \quad S_f(i) = \langle t_i, v_i \rangle
\]
such that $t_i \leq t_{i+1}$ for all $i \in \mathbb{N}$, with $t_i \in T$ and $v_i \in D_f$.  \end{definition}
We denote by $\mathcal{S}$ the set of all streams.
    
    
    
    

\begin{definition}[Stream Operator]
Let $S_1, \ldots, S_n$ be input streams, where $n \geq 0$. A stream operator is a function
\[
O: S_1 \times \cdots \times S_n \rightarrow S_{\text{out}}
\] 
mapping input streams to an output stream $S_{\text{out}}$.
\end{definition}
Special cases include the \emph{source} and the \emph{sink} operators. The former has no input streams and produces a stream from a stream source, namely $O_{\text{source}} : \emptyset \rightarrow S_{out}$. 
Conversely, the latter consumes one or more input streams but produces no output stream, namely $O_{\text{sink}} : S_1 \times \cdots \times S_n \rightarrow \emptyset$.
We denote the set of streams operators by $\mathcal{O}$ and by $\mathcal{O}_{source}$,$\mathcal{O}_{sink}\in \mathcal{O}$ respectively the set of source and sink operators.

\begin{definition}[Transformation Pipeline]
Given the set of streams $\mathcal{S}$ and stream operators $\mathcal{O}$, a  transformation pipeline is a directed acyclic graph (DAG)
\[
G = (\mathcal{O}_G, \mathcal{E}_G)
\]
where $\mathcal{O}_G\subseteq \mathcal{O}$ is the set of nodes, consisting of stream operators,
and $\mathcal{E}_G \subseteq \mathcal{O}_G \times \mathcal{O}_G \times \mathcal{S}$ is the set of directed edges representing dataflows between operators.
Each edge $e = (o_i, o_j, S_k) \in \mathcal{E}_G$ represents the stream $S_k$ flowing from $o_i\in \mathcal{O}_G$ to $o_j\in \mathcal{O}_G$.
\end{definition}
The graph is \emph{acyclic}: there exists no sequence of edges 
$(e_1, e_2, \ldots, e_k)$ with $k\geq 1$, such that each edge $e_i = (o_i, o_{i+1}, S_i) \in \mathcal{E}_G$, and $o_{k+1} = o_1$.
We denote by $\mathsf{input_G}$ the set of input sources for the pipeline, namely $\mathsf{input}_G=\{S_k\in \mathcal{S} \mid \forall (o_i,o_j,S_k)\in \mathcal{E}_G, o_i
\in \mathcal{O}_{source}  \}$, conversely by $\mathsf{output}_G$ its output streams.

We denote by $R$ the set of roles, i.e.  functions within the organization.
We define the function $Right:R\rightarrow 2^\mathcal{S}$ associating each role $r\in R$ to the set of streams that the role is authorized to access.

Set $W$ denotes activities/processes.
An agent, being it an employee or a smart object (e.g., the HVAC system), is defined as follows.
\begin{definition}[Agent]
Let $I_A$ be a countable set of unique identifiers, $R$ a set of roles, $W$ a set of activities/processes, 
an agent $a$ is a tuple
\[
a = \langle id_a, r_a, \lambda_a, w_a, P_a  \rangle
\]
where $id_a \in I_A$, $r_a \in R$, $\lambda_a \in \Lambda$, $w_a \in W$, and the set $P_a$ encodes agent-specific configurations or constraints.
\end{definition}

We denote by $A$ the set of all agents. A \emph{context} for an agent is defined as the tuple of its dynamic properties, namely 
 $\langle r_a,\lambda_a, w_a \rangle$.


Given two agents $a_1,a_2\in A$, a collaboration/delegation relation from an originator agent $a_1$ to a receiver agent $a_2$ is a relation that extends the rights of $a_2$ with some rights of $a_1$ during the performance of an activity $w\in W$. 
We define the set of all valid collaborations as:
$C = \{ (a_1, a_2, w, S_{extra}) \in (A \times A) \times W \times 2^\mathcal{S} \mid S_{extra} \subseteq Right(r_{a_1}) \}$, where $(a_1, a_2, w, S_{extra})$ represents the collaboration tuple. The condition $S_{extra} \subseteq Right(r_{a_1})$ enforces that the transferred rights regard a subset of the streams authorized for the originator’s role.
To make an example, a collaboration from Marco to Anne can be written as $\langle$Marco, Anne, Emergency Line Restoration, $\{S_{Line1}\}\rangle$ $\in C$, where $S_{Line1}$ includes all sensors in the production line 1.


Hereby, we discuss a contextual, operational model of access, which evaluates whether an agent can interact with particular elements of the system (streams, pipelines), based on its context.
Accessibility is assessed both on the basis of the role itself (static rights), and also on dynamic relations (delegation/collaboration) among agents.

A stream is accessible by an agent either (1) if the agent's role possesses a right on the stream or (2) the agent is the recipient of a delegation/collaboration relation on the stream (limited while performing a particular activity/process) which grants the role.

\begin{definition}[Stream accessibility]
\label{def:stream_accessibility}
Let $S\in \mathcal{S}$ be a stream, $a \in A$ be an agent with role $r_a$. 
The predicate $\mathsf{accessible}(a, S)$ holds if and only if $S\in Right(r_a)$ $\vee$  $S \in \{ S_{extra} | \langle a_x,a,w,S_{extra}\rangle \in C\}$.   
\end{definition}


The concept of stream accessibility can be extended to a stream transformation pipeline: the output streams for a pipeline are accessible to an agent if all its input streams are accessible.

\begin{definition}[Pipeline accessibility]
\label{def:pipeline_accessibility}
Let $G$ be a pi\-pe\-line, let $\mathsf{output}_G$ be its output streams and let $a \in A$ be an agent. The following implication is defined:
 $(\forall S_i \in \mathsf{input}_G$ $\mathsf{accessible(a,S_i)})$ $\rightarrow$
 $(\forall S_o \in \mathsf{output}_G$
 $\mathsf{accessible(a,S_o)})$.
\end{definition}



\section{Knowledge Graph representation}
\label{sec:metadata}
Having introduced the formal constructs underlying the model, we now describe how these concepts are implemented using a Knowledge Graph-based infrastructure, enabling semantic interoperability, extensibility, and precise reasoning over data streams and their transformations.
In this model, metadata about sensors, agents, roles, rights, preferences, streams, operations, and transformation pipelines is structured as a labeled, directed RDF graph. 
The knowledge graph serves as a central metadata backbone, providing, unlike flat key-value metadata models, a flexible, schema-rich abstraction that captures not only the structural properties of components but also their semantic interconnections. 

To support modular reasoning and maintain separation of concerns, we organize the knowledge into three interlinked Knowledge Graphs: a \emph{Domain graph} (Subsection \ref{subsec:semioe}) encodes information on devices, agents, locations, roles and rights, a \emph{Stream gathering graph} (Subsection \ref{subsec:technical-kg}) includes metadata on streams, while the \emph{Stream transformation graph} (Subsection \ref{subsec:pipeline-kg}) encodes knowledge on stream pipe\-lines.

\subsection{Domain graph}
\label{subsec:semioe}
\begin{figure}[t]
    \centering
    \includegraphics[width=\linewidth]{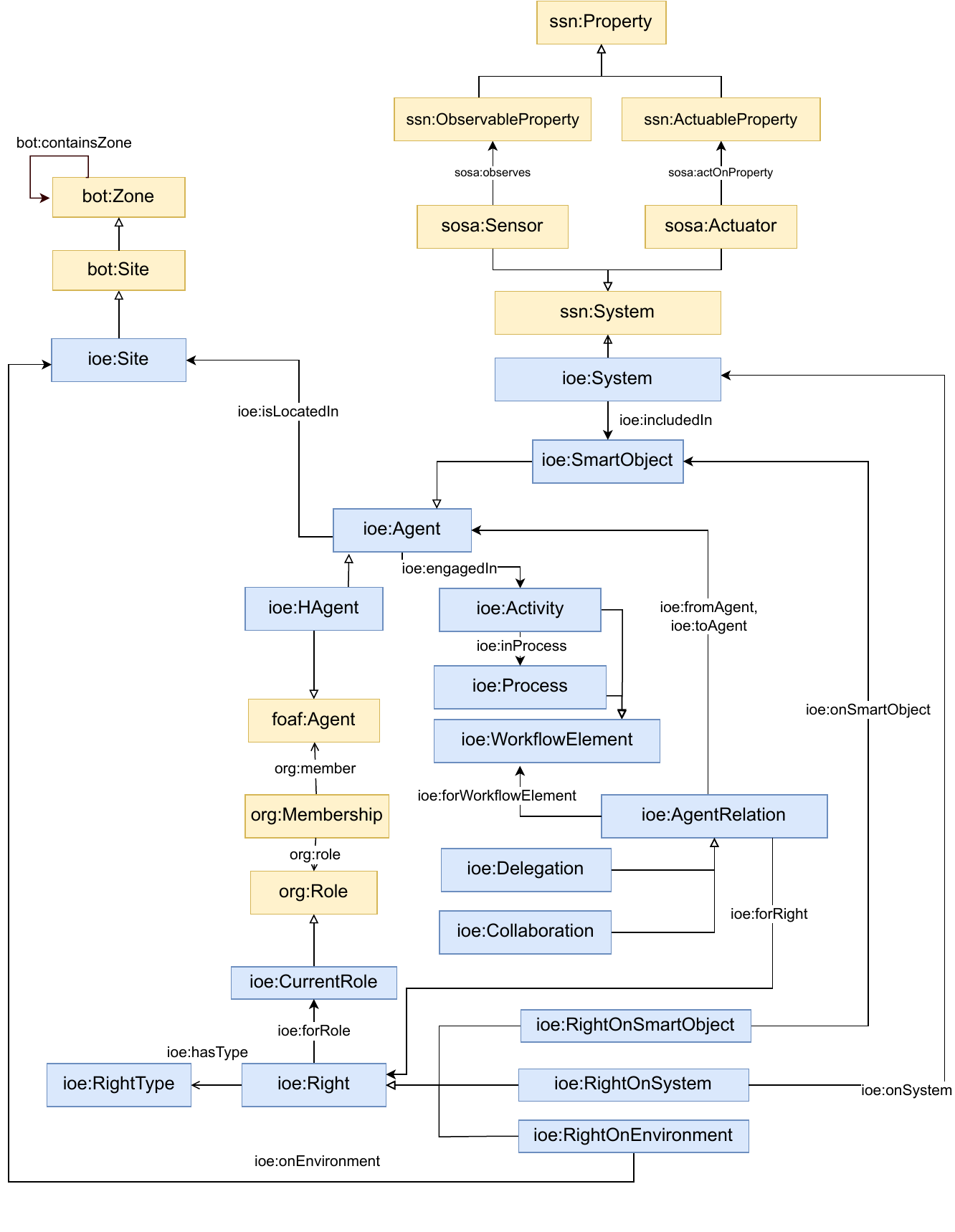}
    \caption{An overview of the SemIoE ontology module (in yellow: external classes).}
    \label{fig:semioe}
   
\end{figure}

The domain Knowledge Graph is built by referring to the SemIoE ontology \cite{arazzi2024semioe} as its schema. SemIoE is a lightweight OWL2 ontology designed to provide a structured and standardized framework to describe entities and their relationships for an Industry 5.0 scenario. The ontology offers a semantic layer representing agents, systems, environments, processes, rights and preferences along with their interconnections within an IoE network. This helps to enhance the semantic understanding of IoE environments, fostering interoperability across heterogeneous IoE components. 
As shown in the diagram in Figure \ref{fig:semioe}, the ontology reuses and integrates several external modules that cover specialized aspects of the IoE landscape. Notably, SemIoE incorporates the W3C Semantic Sensor Network (SSN) ontology \cite{compton2012ssn}, which serves as the foundation for defining technical characteristics and functionalities of sensors and actuators (prefix \emph{ssn} in figure).

Hereby, we summarize the classes most relevant for the purpose of this work in the following (with prefix \emph{ioe} in figure).

The class \emph{Site} represents locations within the organization boundaries and corresponds to the set $\Lambda$ in the model. The class extends \emph{bot:Site} from the Building Topology Ontology \cite{rasmussen2021bot} (prefix \emph{bot} in figure), which in turn is a specialization of \emph{bot:Zone} and inherits the transitive relation \emph{bot:containsZone}. This last enables to define a partial order between specific sites (e.g., Facility A $contains$ Area A2 which $contains$ Production line A2/1).


An \emph{Agent} represents either the class of \emph{HAgent}s, i.e. human agents, or \emph{Smart Object}s. In the model it is denoted by $A$. It is located in a \emph{Site} and can be involved in an \emph{Activity}, which in turn is part of a \emph{Process} (both are specialization of \emph{WorkflowElement}).
A Smart Object, on the other hand, is composed by one or more \emph{Systems}, e.g., a $CO_2$ sensor or a damper actuator for ventilation, which correspond to the set $\mathcal{F}$ in the model. The relation between a system and a smart object is represented through the property \emph{includedIn}. According to the SSN ontology, each system is characterized by a set of technical \emph{Properties}, which correspond to system capabilities $M$. 

\begin{figure}[ht]
    \centering
    \includegraphics[width=\linewidth]{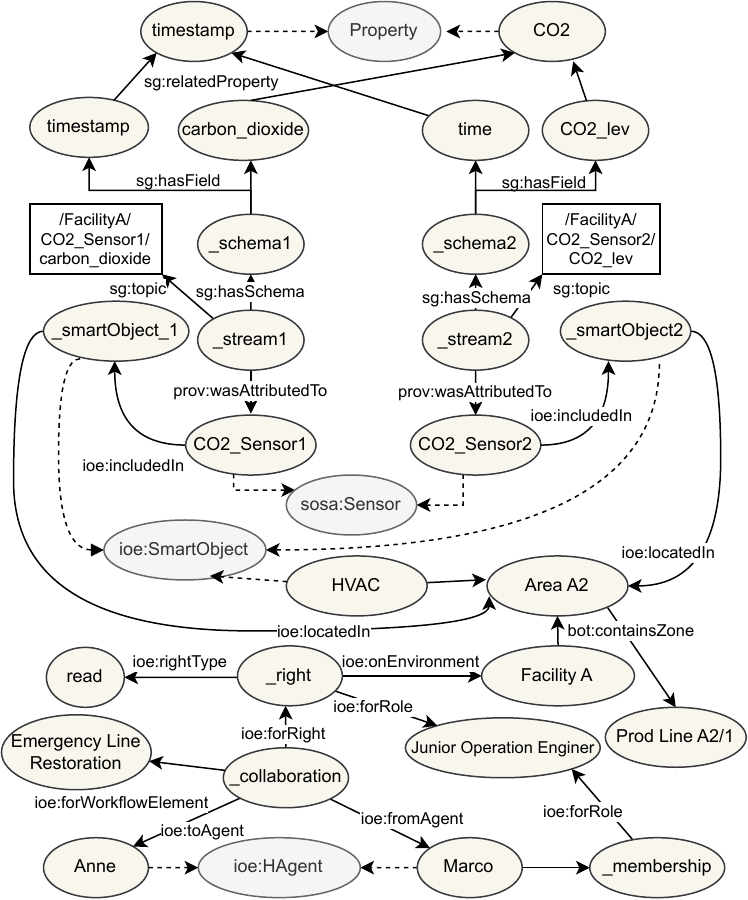}
    \caption{Fragment of the domain/stream gathering KGs.}
\label{fig:example_kg}
\end{figure}

Finally, an Agent can access a number of systems, based on its \emph{Role}. This is represented through a set of rights enabling read/right access to specific systems (\emph{Right\-On\-System}), or to an entire smart object (\emph{Right\-On\-Smart\-Object}), and systems within, or to all systems within an environment (\emph{Right\-On\-Environment}). As such, it allows to represent the set of accessible systems $S_{a_i}$ for an agent $a_i$. Role and rights for an agent can dynamically change with \emph{Collaboration}/\emph{Delegation} relations. These link the granting agent to the receiving agent and are defined by a specific start and end time. Furthermore, these relationships can be bound to a \emph{WorkflowElement}, ranging from a single activity to a full process. An example fragment of the domain graph is shown in Figure \ref{fig:example_kg}.

The SemIoE ontology includes further classes and relations, e.g., to represent preferences, which are not fully reported here. The full specification of the ontology is available at \url{https://w3id.org/semioe}.

\subsection{Stream gathering graph}
\label{subsec:technical-kg}
To model the structure and behavior of streaming data within the IoE environment, we introduce a Stream Gathering graph, which focuses on the representation of real-time data streams, their associated schemas, and processing specifications. This graph's schema is formalized using a dedicated minimal OWL2 ontological module that complements SemIoE by capturing the technical configuration and flow of data within smart systems. It is inspired by IoT-Streams \cite{iot-stream} and similar lightweight stream ontologies, although focusing on operational features of the stream, including the message schema, its topic (namely the name of the logical channel used to categorize messages in the platform), and specifications for monitoring/querying.

\begin{figure}[ht]
    \centering
    \includegraphics[width=\linewidth]{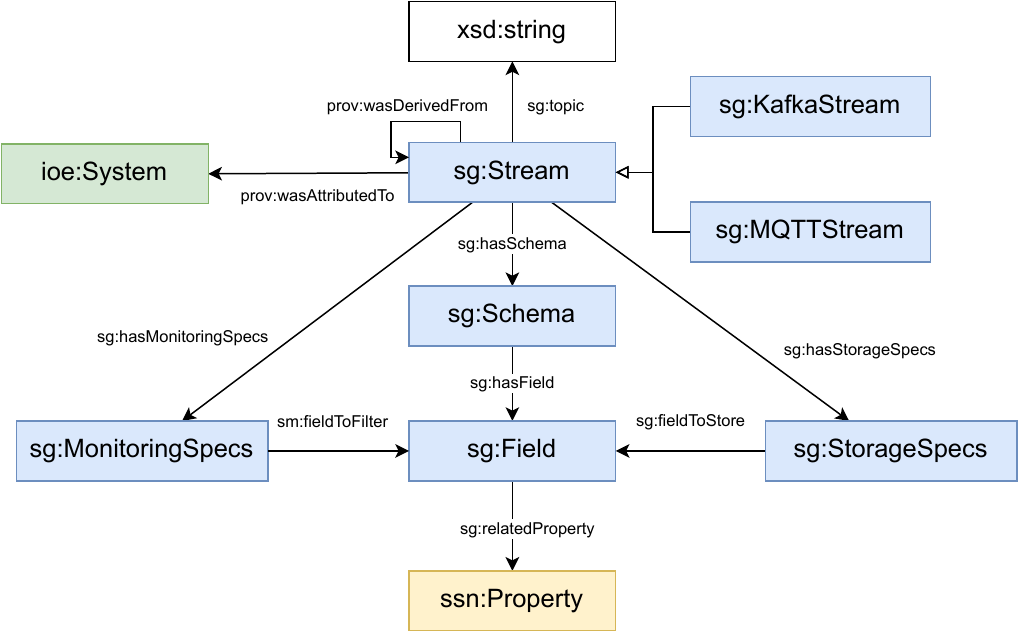}
    \caption{Stream gathering ontological module (in yellow: external classes. In green: class from SemIoE).}
    \label{fig:stregath_ontology}
\end{figure}

In Figure \ref{fig:stregath_ontology}, the main classes and relations are sketched  (with prefix \emph{sg}). The core of the graph lies the class \emph{Stream}, which abstractly represents any flow of data, such as monitored values from an \emph{ioe:System}. The ontology defines two main specializations of this class: \emph{KafkaStream} and \emph{MQTTStream}, which correspond to specific messaging protocols commonly adopted in industrial and IoT scenarios, although further subclasses can be defined. Each stream is identified by a \emph{topic} and is associated with a \emph{Schema} that defines its structural blueprint.

The \emph{Schema} class groups one or more \emph{Fields}, each of which denotes a specific element in the stream’s payload (e.g., temperature, status, or timestamp). Every field is characterized by its \emph{fieldPath} (defining its location in the message structure), \emph{fieldType} (e.g., integer, float, string), and a semantic linkage to a \emph{sosa:Property}, which allows integration with the broader semantic model of system capabilities defined in SemIoE.

In order to support both data processing and storage operations, the model includes two additional classes: \emph{MonitoringSpecs} and \emph{StorageSpecs}. These specify, respectively, which fields in a stream are relevant for monitoring (e.g., for triggering alerts or control actions), and which should be persisted for historical analysis or compliance purposes. The \emph{StorageSpecs} class also contains metadata about the destination storage system, such as the DBMS type, the database name, and the target table/collection.

\subsection{Stream transformation graph}
\label{subsec:pipeline-kg}
\begin{figure}
    \centering
    \includegraphics[width=\linewidth]{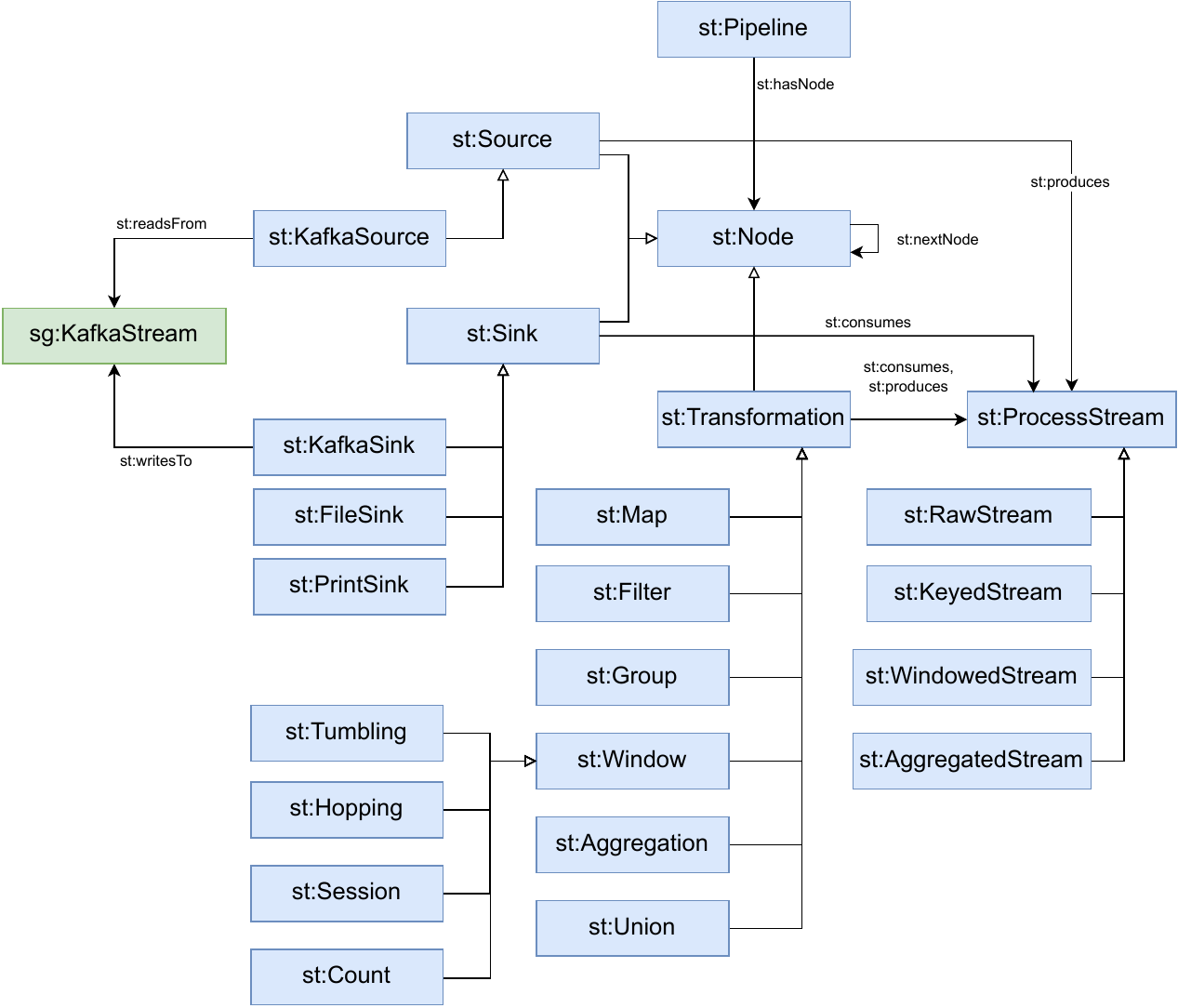}
    \caption{Stream transformation ontological module (in green: class from the Stream gathering module).}
    \label{fig:StreDag_ontology}
\end{figure}

The Stream transformation graph is aimed to represent the  transformations applied to streams. Its schema is defined through a lightweight OWL2 ontological module, whose main classes and relations are represented in Figure \ref{fig:StreDag_ontology} (with prefix  \emph{st}).

Its core includes the class \emph{Pipeline}, which acts as a container for a sequence of interconnected \emph{Node}s. Each node models a logical stage in the data flow and is connected to the next via the property \emph{hasNext}, forming a directed acyclic graph (DAG) structure. The ontology includes  three specializations of nodes: \emph{Source}, \emph{Transformation}, and \emph{Sink}.

\emph{Source} nodes initiate the stream processing pipeline, e.g., the \emph{KafkaSource} class represents a node that reads from a Kafka topic (linked to a \emph{sg:KafkaStream} from the Stream Gathering graph via \emph{readsFrom}). Each source produces a \emph{ProcessStream}, which is passed along the pipeline.

\emph{Transformation} nodes represent the functional core of the pipeline and include stream operations such as \emph{Map}, \emph{Filter}, \emph{KeyBy}, \emph{Aggregation}, and \emph{Union}. These transformations consume and produce \emph{ProcessStream}s, with the ability to operate on specific fields via the \emph{appliesToField} property. Aggregations specify a function (e.g., sum, avg) through the \emph{aggregationFunction} property. The ontology also models windowing semantics using the \emph{Window} class and its specializations: 
 \emph{Tumbling} is a fixed-size (property \emph{windowDuration}), non-overlapping window,  \emph{Hopping} is a fixed-size and possibly overlapping window (property \emph{windowHop} determines the interval), 
\emph{Session} is a window dynamically sized based on inactivity gaps, and 
 \emph{Count} is a window collecting a  number of events (property \emph{size}), and is not based on time.

ProcessStreams are distinguished in various types, e.g., \emph{RawStream}, \emph{KeyedStream}, and \emph{WindowedStream}, which are aimed to support validation of transformation pipelines. Each transformation is intended to accept an input streams of specific types, enabling type-safe composition and reasoning about pipeline correctness, which is however left to future work.

\begin{figure}[ht]
    \centering
    \includegraphics[width=\linewidth]{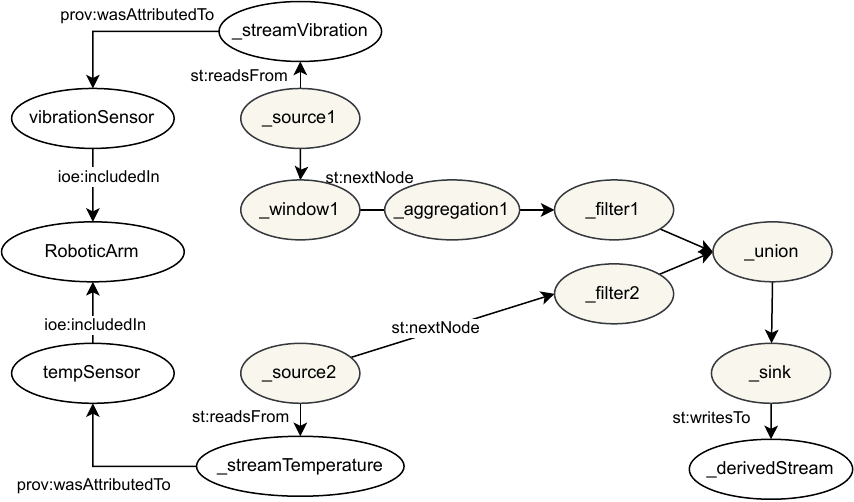}
    \caption{Fragment of stream transformation KG (in white: instances of other graphs).}
    \label{fig:pipeline}
\end{figure}

At the end of the pipeline, \emph{Sink} nodes store or emit the transformed data. Sinks include \emph{PrintSink}, \emph{FileSink}, and \emph{KafkaSink} as subclasses. The latter can write the output stream back to a new Kafka topic via the \emph{writesTo} property.
The model enables multi-branch pipelines, where the same stream may be consumed by different stream operators. This enables a flexible, fine-grained control over how data is processed, filtered and routed across parallel branches of the same pipeline.
Figure \ref{fig:pipeline} shows a fragment of the KG modeling a pipeline which reads streams from a high-frequency vibration sensor and a low-frequency temperature sensor mounted on a robotic arm. The pipeline applies several transformations, including windowed aggregation of the vibration data and filtering to identify local anomalies. Finally, the processed streams are fused into a single derived stream.

\section{Stream Management Platform}
\label{sec:framework}

\begin{figure}[ht]
    \centering
\includegraphics[width=\linewidth]{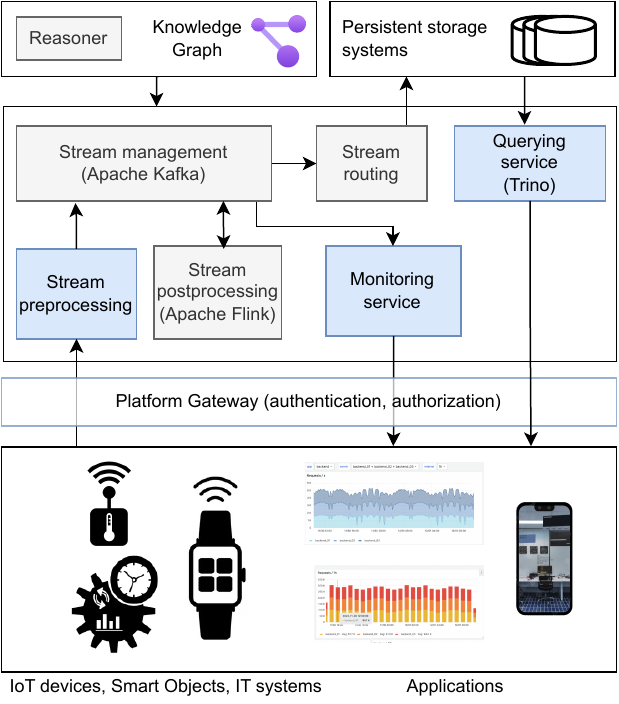}
 \caption{Platform architecture}
    \label{fig:architecture}
\end{figure}

This section provides a detailed explanation of platform architecture, which relies on the Knowledge Graph model to provide context-aware stream management and support to analysis.
The architecture is structured in layers as depicted in Figure \ref{fig:architecture}.
 The data layer consists of the Knowledge Graph for resource metadata, supported by a reasoner, and storage systems for data streams. 
 The data management layer, in turn, adopts a modular design that is organized into a set of microservices, each responsible for handling distinct aspects of the data stream processing pipeline:
%
  \emph{data gathering} (Subsection \ref{subsec:management}), which supports stream acquisition, 
  \emph{stream transformation} (Subsection \ref{subsec:transformation}), 
   \emph{semantic data monitoring} for real-time stream consumption (Subsection \ref{subsec:monitoring}), \emph{data querying} for data extraction from storage systems (Subsection \ref{subsec:querying}). Finally, the \emph{authentication and authorization} layer (Subsection \ref{subsec:authentication}) serves as a gateway to the rest of the platform. The platform implementation is available at the project's GitHub repository\footnote{https://github.com/Homey-Prin22/framework}.

\subsection{Data gathering}
\label{subsec:management}
Real-time data streams originate from various sources, including IoT sensors, wearable devices, and smart objects, in heterogeneous formats and communication protocols. Data streams may be transmitted directly to the platform or routed through local gateway nodes that aggregate data before forwarding to message brokers, like Mosquitto: for instance, temperature and humidity sensors typically publish MQTT data to gateways, while wearable devices often transmit physiological data via Bluetooth Low Energy (BLE) to local aggregators (e.g. smartphones), which then forward data via HTTP. 
A dedicated preprocessing layer performs domain-specific transformations (e.g., pre-filtering, decompressing or decrypting) before republishing to new Kafka topics. 

At the core, an Apache Kafka\footnote{https://kafka.apache.org/} cluster serves as the primary real-time data stream processing platform. Kafka is a distributed, fault-tolerant, and high-throughput platform that organizes data into topics and partitions (see Subsection \ref{subsec:monitoring} for details), enabling parallel consumption and horizontal scalability through a publish-subscribe pa\-ra\-digm. Producers send messages to topics, without knowledge of consumers, while multiple consumers can independently subscribe to topics. This ensures parallel data consumption, loose coupling between producers and consumers, system evolution, heterogeneous data stream integration, along with scalability and fault tolerance. 



\subsection{Stream transformation}
\label{subsec:transformation}
The architecture integrates Apache Flink\footnote{https://flink.apache.org/} within the stream post-processing service to support structured, persistent stream processing workflows. Flink is a distributed framework designed for stateful computations with event time semantics, complex windowing operations, and sophisticated late data handling. Its architecture is optimized for low latency and high throughput performance. 
Flink applications are structured as jobs that define a Directed Acyclic Graph, with sources, processing pipeline, and sinks. The model supports a broad set of stream processing patterns, including event-time windows and custom aggregations, enabling real-time transformations such as sliding-window aggregation, filtering, and enrichment over data streams. The framework supports monolithic persistent pipelines and dynamic on-demand job instantiation. For instance, when a technician requires aggregated sensor data over specific time windows, the system can dynamically activate a Flink job configured to consume from corresponding Kafka topics, perform the specified aggregation logic in real-time, and publish results to designated output topics.
Job orchestration is driven by semantic metadata in the stream transformation Knowledge Graph. 

As such, every stored transformation pipeline is a declarative specification of a DAG (see Figure \ref{fig:pipeline} as an example), from which  
Flink jobs are automatically instantiated, defining: (i) source ingestion, with timestamp extraction and watermark; (ii) transformation stage, with filtering and mappings; (iii) optional windowing and aggregation stage; and (iv) output routing to sinks (e.g., printing on a console or publishing on a Kafka topic). This approach allows reconfigurable stream processing pipelines through metadata, without modifying the core application logic. 



\subsection{Data Monitoring}
\label{subsec:monitoring}
When dealing with publisher-subscriber architectures, a key concept is that of \emph{topics}, i.e., named logical channel used to organize and route messages.
In standard monitoring systems topics are often hierarchically structured, e.g., a topic like \texttt{Area\_A2/CO2\_sensor1/carbon\_dioxide} includes three dimensions specifying their order within the topic string, namely site, device and measure.
The utilization of wildcards (in MQTT brokers) or regex (in Kafka) provide a degree of subscription flexibility, e.g., \texttt{Area\_A2\textbackslash..*} expresses any stream coming from the site.

However, this purely syntactic approach introduces significant architectural challenges.
The most prominent limitation is the inherent structural rigidity of the system.
Different naming conventions among device manufacturers complicate integrated data management, e.g., a sensor can refer to CO2 as ``carbon\_dioxide'' while another as ``CO2\_lev''.
Furthermore, as the topic hierarchy is fixed, the introduction of new dimensions, such as a ``facility'' or ``process'' level, would require a comprehensive revision of the naming convention. This results in tight coupling, as client applications must be manually updated to align with the modified schema.
As system complexity increases, maintaining consistency across topic hierarchies can become increasingly error-prone, ultimately undermining the scalability and reliability of the messaging infrastructure. 


Semantic monitoring extends the traditional monitoring paradigm, as it decouples data stream access from fixed topic naming, by leveraging the semantic relationships stored in the KG where each sensor, and related stream, is completely contextualized.
In this way, it is possible to find sensors by what they measure or their properties, rather than just their topic names.
Requests are formulated as a conjunctive set of constraints, where each constraint can specify one or more instance of classes \emph{Property}, \emph{Site} and one or more URIs of \emph{Sensor}. The service uses these constraints to compose a SPARQL query that is executed on the Knowledge Graph, possibly exploiting logical reasoning to infer additional knowledge. 
To make an example (see Figure \ref{fig:example_kg} for reference), let us assume the smart object \emph{HVAC} needs to monitor $CO_2$ in Area A2. The constraints $[\{Property:[CO2]\}, \{Site:[Area\_A2]\}]$ aim (1) to identify all relevant devices in the Knowledge Graph, i.e. \emph{CO2\_Sensor1} and \emph{CO2\_Sensor2} which provide a \emph{CO2} measurement (indipendently on how they were named in the stream schema) and are located in \emph{Area\_A2}, their data schema and topic names.
The query filters out streams that are not accessible to the agent that formulated the request, hence providing a context-based access control based on the current rights (see also Section \ref{sec:reasoning}).

\begin{figure}[ht!]
    \centering
    \includegraphics[width=\linewidth]{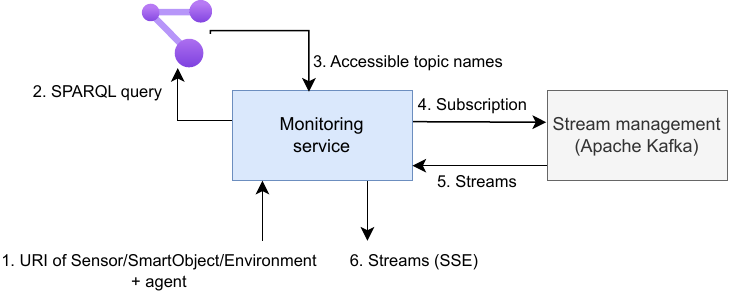}
    \caption{Interaction workflow of the monitoring service.}
    \label{fig:monitoring_service}
\end{figure}

The workflow, shown in Figure \ref{fig:monitoring_service}, starts when the agent formulates the monitoring request (1), which is translated into a SPARQL query (2). The service retrieves topic names accessible by the agent (3), and  dynamically subscribes to them (4). Incoming data are forwarded to the client via Server-Sent Events (SSE), enabling dynamic and context-aware monitoring (5-6).
The combination of scalable stream handling, secure access control, and semantic enrichment makes the monitoring layer a pivotal architectural element, covering the gap between raw data flows and high-level context-aware applications, supporting the shift toward a semantic-driven approach.

\subsection{Data Storage and querying}
\label{subsec:querying}

Persistent storage enables to store both raw and derived data streams into appropriate Da\-ta\-base Management Systems (DBMSs), selected based on data type, data structure, schema complexity, and use cases. Our architecture integrates multiple DBMS technologies (e.g., PostgreSQL, MongoDB and MySQL):  time-series da\-ta\-ba\-ses for high-frequency sensor logs, document-oriented stores for semi-structured data, and relational databases for highly structured records. 
Metadata concerning storage configurations and routing rules are included in Stream gathering KG (as \emph{StorageSpecs} specifying the DBMS, the dataset name, the table/collection, and which fields are to be stored). For each data stream, the \emph{Routing} service subscribes to Kafka topic(s) and writes data to the appropriate DBMS. This modular approach supports flexible persistence. 

However, data fragmentation across multiple DBMSs introduces a critical challenge to perform complex time-based queries. A na{\"i}ve approach might be querying each database independently and merging results at the application layer, but it proves neither efficient nor scalable. 
To address this problem, the proposed architecture integrates a dedicated \emph{Querying} module that abstracts the complexity of accessing multiple storage systems through a semantic-based querying mechanism and federated queries. 
Clients interact via a web interface, specifying mandatory parameters (e.g., sensor name), and optional parameters, including time windows, aggregation functions (average, min, max), sorting, and filtering conditions. 
\begin{figure}[ht]
    \centering
    \includegraphics[width=\linewidth]{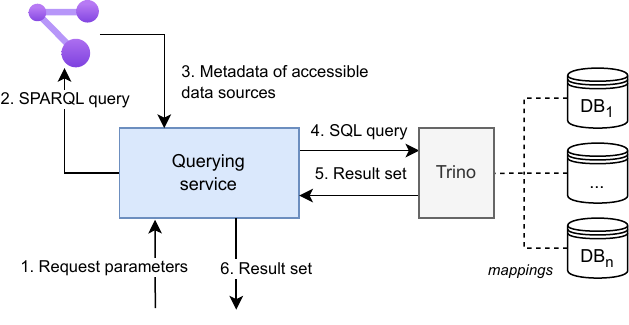}
    \caption{Interaction workflow of the Querying service.}
    \label{fig:querying_api}
\end{figure}
The module provides uniform data retrieval across heterogeneous storage systems through a semantic-based mechanism, shown in Figure \ref{fig:querying_api}. The workflow proceeds as follows: the service receives client requests (1) and semantically interprets parameters, leveraging the Stream gathering KG to identify the relevant streams accessible by the agent (according to the agent context, and possible collaboration/delegation relations), data sources, and the associated storage systems 
(2-3). For each identified accessible stream, the API constructs SQL subqueries, retrieving required measurements from corresponding databases, composing multiple subqueries into a unified SQL statement when needed. The query is submitted to Trino (4), a distributed SQL query engine for federated queries across different backend systems. 
Trino's workers retrieve data from heterogeneous sources (e.g., PostgreSQL, MySQL, MongoDB) to compute the final result set (5), and returns it to clients (6).

\subsection{Authentication and Authorization}
\label{subsec:authentication}


An Authentication API is implemented as the single exclusive external entry point, acting as a gateway to all internal services. External requests are intercepted and client identities are verified through JWT (JSON-Web Token) based authentication. 
Upon successful login, the API generates access and refresh tokens containing user identity information and validity periods, so the client is successfully authenticated and the client is allowed to proceed with requests through the gateway, e.g., to access  Monitoring or Querying services.
On the one hand, service-level authorization is performed to check whether the client is authorized to access the specified service. 
User credentials are securely stored using Flask-Bcrypt hashing and salting techniques to prevent plain text password retrieval. 
For real-time performance and low-latency responses, the API employs Redis for fast in-memory transient information management. 

On the other hand, as discussed in previous subsections, stream-level authorization is evaluated at each service, following a Role-Based Access Control (RBAC) model: access rights are dynamically determined based on the agent's  operational context retrieved from the KG. The next section discusses how this is realized through semantic reasoning techniques. 

\section{Context-aware reasoning services}
\label{sec:reasoning}
This section discusses reasoning functionalities exploiting the graph to derive context-aware knowledge supporting platform services, including stream monitoring and querying, and related authorization procedures. Reasoning services enable, among other capabilities, the following functions:
\begin{itemize}
    \item check if a sensor/stream is accessible by an agent, based on the current role; 
    \item check if the output stream from a transformation pipeline is accessible by an agent, based on the current role; 
    \item retrieve all accessible sensors, and related topics, co-located with an agent.
\end{itemize}

Since part of the knowledge is dynamic and can change at any time (specifically the user context, including the current activity and its location), knowledge extraction is decoupled in two separate steps. At first, (1) reasoning on  static knowledge, including the industrial topology, deployment of devices and their technical specifications, is performed through OWL axioms (e.g., transitivity of ObjectProperies) and SWRL rules \cite{horrocks2004swrl}. These last are used to  express complex conditional logic and domain-specific constraints that cannot be captured by ontology axioms alone, enabling rule-based inference beyond standard description logic reasoning.
As a result, inferred triples through deductive closure are then materialized in the graph. 
Then, at run-time, (2) SPARQL queries are executed on the enriched graph to derive the final results.

On the one hand, transitive properties, such as \emph{bot:con\-tains\-Zone} are processed in order to compute their transitive closure, thereby enabling the inference of indirect containment relationships across hierarchical structures (e.g., if a \emph{facility} contains an \emph{area}, which, in turn, contains a \emph{production line}, then this last is declared to be contained in the \emph{facility}).
On the other hand, SWRL rules are defined for various goals, including associating each role’s rights (e.g., on an environment) with the specific  sensors that are actually accessible, as shown in Listing \ref{listing:swrl}.

\begin{lstlisting}[caption=SWRL rule to derive access rights on sensors based on location containment., label=listing:swrl]
ioe:onEnvironment(?r,?env) ^ 
ioe:includedIn(?s,?sm) ^ 
ioe:isLocatedIn(?sm,?env) 
-> ioe:onSensor(?r,?s)
\end{lstlisting}

A similar rule allows to derive accessible sensors from rights on smart objects. 

Please note that SWRL cannot express some conditions. In particular, non-monotonic, closed-world, and universally quantified reasoning is not supported by the Horn-logic foundation of SWRL. Therefore, we express such conditions directly using SPARQL queries (with FILTER NOT EXISTS or MINUS constructs). Alternative solutions for enforcing this class of constraints include SHACL or a custom rule engine.

In Listing \ref{listing:query}, a SPARQL query is used to extract streams accessible to an agent with a given role  (both those generated by sensors for which the role grants read permissions and those derived through transformation pipelines).

\begin{lstlisting}[caption=SPARQL query to extract streams accessible to a given <role>., label=listing:query, basicstyle=\ttfamily\small]
SELECT DISTINCT ?stream
WHERE {
VALUES ?role {<role>}
  { ?right ioe:forRole ?role .
   ?right ioe:onSystem ?s .
   ?stream prov:wasAttributedTo ?s .}
   
  UNION 
  
  {?stream prov:wasDerivedFrom ?b .
   FILTER NOT EXISTS {
    ?stream prov:wasDerivedFrom ?b2 .
    ?b2 prov:wasAttributedTo ?s2 .
    FILTER NOT EXISTS {
      ?right ioe:forRole ?role .
      ?right ioe:onSystem ?s2 .}
    }
   }
}
\end{lstlisting}


In this case, the use of the \texttt{MINUS} keyword ensures that streams derived from pipelines are included only if all their base streams are accessible to the agent’s role, which corresponds to a universal quantification over the base streams. To constrain sensors to be co-located with the agent, it is sufficient to add the pattern  \texttt{<agent> ioe:isLocatedIn ?p} and \texttt{?s ioe:isLocatedIn ?p}.

In Listing \ref{listing:query2}, a SPARQL extracts the streams accessible to an <agent> performing an <activity> based on contextual collaboration/delegation relations.

\begin{lstlisting}[caption=SPARQL query to extract streams accessible to an agent performing an activity based on collaboration/delegation., label=listing:query2, basicstyle=\ttfamily\small]
SELECT DISTINCT ?stream
WHERE {
VALUES (?agent ?w) {<agent><activity>}
  { ?c a ioe:AgentRelation;
       ioe:forWorkflowElement ?w;
       ioe:toAgent ?agent;
       ioe:forRight ?r.
    ?r ioe:onSensor ?s.
    ?stream prov:wasAttributedTo ?s.}
}
\end{lstlisting}

  

\section{Evaluation}
\label{sec:evaluation}
This section presents an evaluation of the platform, focusing on execution performance for queries over Knowledge Graphs (Subsection \ref{subsec:eval_KG}), response latency for the monitoring (Section \ref{subsec:eval_monitoring}) and the querying services (Subsection \ref{subsec:eval_querying}), and the end-to-end system performance (Subsection \ref{subsec:eval_platform}).
The experiments were conducted on a commodity system equipped with four 2.30 GHz CPU cores and 16 GB of RAM, running Rocky Linux 9.4. 
REST services were implemented in Python using Flask. 
Details on the specific software versions and Docker instances are available at the project repository.
For all experiments, we assume synthetic sensors sending JSON messages with an average size of 125 bytes to the MQTT broker. Messages are handled by the preprocessing service which directly republishes the stream to Kafka.

\subsection{Knowledge Graph}
\label{subsec:eval_KG}
This evaluation aims to assess query execution time on the Knowledge Graph and  how it evolves as the size of the graph increases, using a set of representative query patterns reported in Table \ref{tab:evaluation_queries}, which correspond to query needed to support monitoring and querying (please note that Q4 is reported in Listing \ref{listing:query}). In particular, the most frequently issued queries are Q1 and Q3, needed to access real-time monitoring, and Q5, used whenever the agent's location changes. 

\begin{table}[h]
\resizebox{.5\textwidth}{!}{
\begin{tabular}{|l|l|}
\hline
Q1) & \begin{tabular}[c]{@{}l@{}}Retrieve the topic of the stream generated  by a \\ \textless{}sensor\textgreater{}\end{tabular}                                                      \\ \hline
Q2) & \begin{tabular}[c]{@{}l@{}}Identify all sensors that generate a stream \\ containing an attribute mapped to a specified \\ global \textless{}property\textgreater{}\end{tabular} \\ \hline
Q3) & \begin{tabular}[c]{@{}l@{}}Verify whether a <stream>  is accessible  by an \\ agent's \textless{}role\textgreater{}\end{tabular}                             \\ \hline
Q4) & Retrieve all streams accessible to a  \textless{}role\textgreater{}                                                                                                              \\ \hline
Q5) & Identify all sensors co-located with the \textless{}agent\textgreater{}                                                                                                          \\ \hline
\end{tabular}}
\caption{SPARQL queries used in the evaluation.}
\label{tab:evaluation_queries}
\end{table}

\begin{figure}[ht]
\centering
\resizebox{\linewidth}{!}{
\begin{tikzpicture}
\begin{axis}[
    width=0.47\textwidth,
    height=7cm,
    xlabel={KG size [\# triples]},
    ylabel={Execution time [ms]},
    xtick=data,
    xticklabel style={rotate=45, anchor=east},
    xmode=log,
    log basis x=10,
    ymode=log,
    log basis y=10,
    xtick={8181,57711,522567,5083016},
    xticklabels={8k, 58k, 523k, 5.1M},
    legend style={
        at={(0.02,0.98)},          
        anchor=north west,
        draw=none,
        fill=white,  
    },
    grid=major,
]

\addplot[
    color=blue,
    mark=*,
    error bars/.cd,
    y dir=both,
    y explicit,
]
table[
    x=triples,
    y=Q1,
    y error=stdev_Q1,
    col sep=tab
] {data/graph_fgcs_2025.dat};
\addlegendentry{Q1}

\addplot[
    color=red,
    mark=square*,
    error bars/.cd,
    y dir=both,
    y explicit,
]
table[
    x=triples,
    y=Q2,
    y error=stdev_Q2,
    col sep=tab
] {data/graph_fgcs_2025.dat};
\addlegendentry{Q2}

\addplot[
    color=green!60!black,
    mark=triangle*,
    error bars/.cd,
    y dir=both,
    y explicit,
]
table[
    x=triples,
    y=Q3,
    y error=stdev_Q3,
    col sep=tab
] {data/graph_fgcs_2025.dat};
\addlegendentry{Q3}

\addplot[
    color=purple,
    mark=diamond*,
    error bars/.cd,
    y dir=both,
    y explicit,
]
table[
    x=triples,
    y=Q4,
    y error=stdev_Q4,
    col sep=tab
] {data/graph_fgcs_2025.dat};
\addlegendentry{Q4}

\addplot[
    color=orange,
    mark=star,
    error bars/.cd,
    y dir=both,
    y explicit,
]
table[
    x=triples,
    y=Q5,
    y error=stdev_Q5,
    col sep=tab
] {data/graph_fgcs_2025.dat};
\addlegendentry{Q5}

\end{axis}
\end{tikzpicture}}
\caption{Execution time for queries Q1-Q5 across KGs of increasing size (log scale).}
\label{fig:results_graph}
\end{figure}
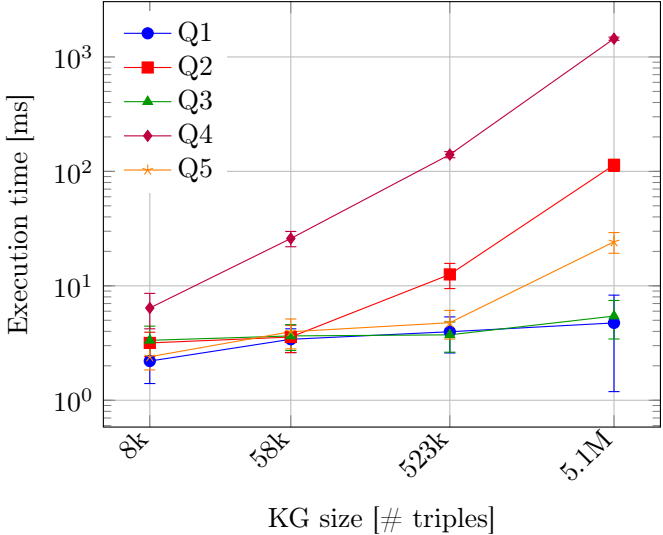

As reported in Table \ref{tab:graph_experiment_setup}, a set of Knowledge Graphs of increasing size have been generated, by varying the number of agents ($\{10,50,250,750\}$), roles ($\{5,15,30\}$), smart objects (${10, 100, 1k, 10k}$) and devices (${100, 1k, 10k, 100k,}$), assuming a number of 10 sensors per smart objects. The fixed parameters comprise 50 global properties, 50 rights per role, and 50 locations. Stream schemas contain between 2 and 5 attributes, the number of pipelines is set to 70\% of the total number of streams, and the average pipeline includes 3 operators.
The resulting graphs have an overall size ranging from 8181 to 5.08 million triples. 
Queries have been executes 10 times each with randomly chosen parameters, and results are averaged.
As shown in Figure \ref{fig:results_graph}, queries Q1, Q3 and Q5, that are also the most frequently issued queries, are executed in less than 10ms in all scenarios. On the other hand, the most complex query results Q4, which is executed only when the agent changes role, takes 1.44s on average in the largest scenario. 

\begin{table}[]
\resizebox{0.5\textwidth}{!}{
\begin{tabular}{c|c|c|c|c|c|}
\cline{2-6}
                         & Agents &  Roles & SO  &  Devices &  Triples \\ \hline
\multicolumn{1}{|c|}{G1} & 10         & 5        & 10     & 100        & 8181       \\ \hline
\multicolumn{1}{|c|}{G2} & 50         & 15       & 100    & 1000       & 57711      \\ \hline
\multicolumn{1}{|c|}{G3} & 250        & 30       & 1000  & 10000      & 522567     \\ \hline
\multicolumn{1}{|c|}{G4} & 750        & 30       & 10000 & 100000     & 5083016    \\ \hline
\end{tabular}}

\caption{Properties of the generated KGs.}
\label{tab:graph_experiment_setup}
\end{table}

\subsection{Monitoring}
\label{subsec:eval_monitoring}
This evaluation aims to assess the response latency of the monitoring service, i.e., the time between sending a  request to the service and receiving the first message in the resulting data stream.
%
The test also aims to quantify the computational overhead introduced by KG queries within the complete monitoring workflow discussed in Subsection \ref{subsec:monitoring}.
For Knowledge Graphs of various sizes (from few thousand to multi-million triples), the experiment involved randomly choosing 10 sensor ids, invoking the service with it as a parameter and averaging the response latency.

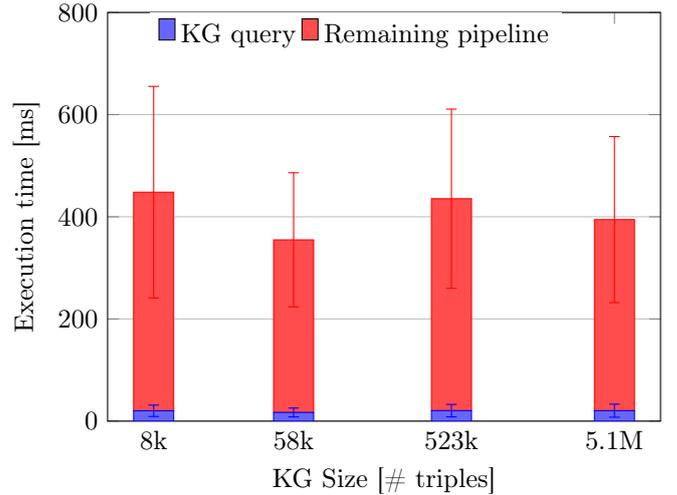
\begin{figure}[ht]
\centering
\begin{tikzpicture}
\begin{axis}[
    ybar stacked,
    width=\linewidth,
    height=7cm,
    bar width=15pt,
    xlabel={KG Size [\# triples]},
    ylabel={Execution time [ms]},
    xmode=log,
    log basis x=10,
    xtick={8181,57711,522567,5083016},
    xticklabels={8k, 58k, 523k, 5.1M},
    ymin=0,
    ymax=800,
    ymajorgrids=true,       
    xmajorgrids=false,
    legend style={
        at={(0.45,1.00)},anchor=north,
        draw=none,
        fill=white,legend columns=-1
    },
]

\addlegendentry{KG query}
\addlegendentry{Remaining pipeline}
\addlegendimage{fill=blue!60, draw=none}
\addlegendimage{fill=red!70, draw=none}

\addplot+[
    fill=blue!60,
    error bars/.cd,
      y dir=both,
      y explicit,
] table[
    x=kg_size,
    y=line1,
    y error=stdev_line1,
    col sep=tab
] {data/monitor.dat};

\addplot+[
    fill=red!70,
    error bars/.cd,
      y dir=both,
      y explicit,
] table[
    x=kg_size,
    y expr=\thisrow{line2}-\thisrow{line1},
    y error=stdev_line2, 
    col sep=tab
] {data/monitor.dat};

\end{axis}
\end{tikzpicture}
\caption{Execution time for monitoring service.}
\label{fig:results_monitoring}
\end{figure}

Results, summarized in Figure \ref{fig:results_monitoring}, show that the overall response latency is below 0.5 seconds in all considered scenarios. This value demonstrates that the system can scale while remaining usable, and is in line with many current IIoT deployments.
In particular, KG query (blue bars in Figure) have a minimal impact, accounting for less than 5\% of total response latency across all evaluated scenarios. The remaining operations (red bars), which consist of topic subscription, data filtering and SSE activation, consistently dominate the execution time profile, representing approximately 95\% or more of the overall duration. 
This highlights that even as the KG scales from 8k to 5.1M triples, the KG query execution time remains remarkably stable, indicating minimal sensitivity to graph size.

\subsection{Querying}
\label{subsec:eval_querying}
This experiment evaluates the response latency of the querying service, i.e., the elapsed time between sending a request and receiving the result set.
The test aims to assess the impact of (1) database size and (2) query complexity on performance. Furthermore, it also aims to evaluate the relative overhead of the Knowledge Graph within the whole querying workflow introduced in Subsection \ref{subsec:querying}.
The experiment was configured as follows: \begin{itemize}
    \item Five datasets including an increasing number of messages generated by a sensor, namely \{3.6k, 36k, 360k, 3.6M,  36M\}. They are implemented using MongoDB collections.
    \item Three common query patterns: selection with ti\-me\-stamp-based filtering (\emph{Filter$_1$}),  with an additional specific field filter (\emph{Filter$_2$}), and selection with time\-stamp-based filtering combined with aggregation (\emph{Aggregation}).
    \item A Knowledge Graph containing 5.08M triples (denoted as G4). 
\end{itemize}
For each dataset size, the service was executed 10 times and results have been averaged.

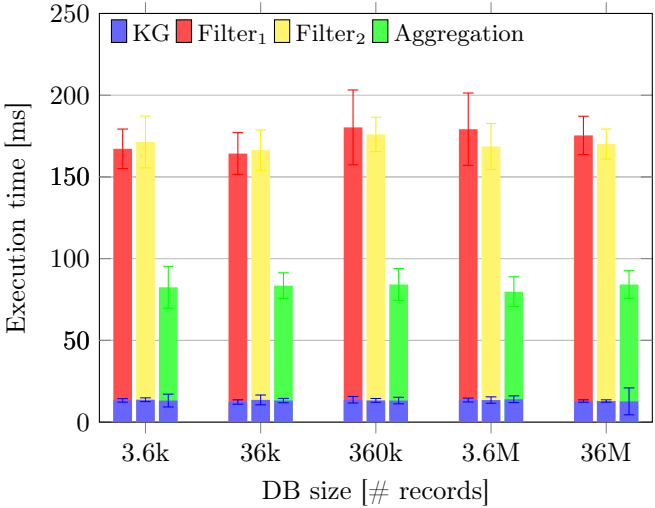
\begin{figure}[ht]
\centering
\begin{tikzpicture}
\begin{axis}[
    ybar,
    bar width=7pt,
    width=\linewidth,
    height=7cm,
    ylabel={Execution time [ms]},
    xlabel={DB size [\# records]},
    xticklabels={3.6k,36k,360k,3.6M,36M},
    xtick={1,2,3,4,5},
    legend style={at={(0.4,1.0)},anchor=north ,draw=none,legend columns=-1,font=\small
    },
    ymin=0,
    ymax=250,
    ymajorgrids=true,       
    xmajorgrids=false,
    extra y ticks={50,150},
    legend image code/.code={
    \draw[#1] (0cm,-0.1cm) rectangle (0.2cm,0.2cm);
}
]

\addlegendentry{KG}
\addlegendentry{Filter$_1$}
\addlegendentry{Filter$_2$}
\addlegendentry{Aggregation}
\addlegendimage{fill=blue!60, draw=none}
\addlegendimage{fill=red!70, draw=none}
\addlegendimage{fill=yellow!70, draw=none}
\addlegendimage{fill=green!70, draw=none}

\addplot+[mark=none, bar shift=-0.3cm,fill=red!70, draw=none,error bars/.cd,
      y dir=both,
      y explicit,error bar style={draw=red}, error mark=|,error mark options={draw=red}]table[x=Test,y=total_end_to_end_ms_mean_q1,y error=total_end_to_end_ms_stdev_q1]{data/query.dat};       

\addplot+[mark=none, bar shift=0cm,fill=yellow!70, draw=none,error bars/.cd,
      y dir=both,
      y explicit,error bar style={draw=yellow},error mark=|, error mark options={draw=yellow}]table[x=Test,y=total_end_to_end_ms_mean_q2,y error=total_end_to_end_ms_stdev_q2]{data/query.dat};
\addplot+[mark=none,bar shift=0.3 cm,fill=green!70, draw=none,error bars/.cd,
      y dir=both,
      y explicit,,error bar style={draw=green}, error mark=|,error mark options={draw=green}]table[x=Test,y=total_end_to_end_ms_mean_q3,y error=total_end_to_end_ms_stdev_q3]{data/query.dat};

\addplot+[bar shift=-0.3cm,fill=blue!60, draw=none,error bars/.cd,
      y dir=both,
      y explicit,error bar style={draw=blue},error mark=|, error mark options={draw=blue}]
table[x=Test,y=getSensorMetadataByID_ms_mean_q1,y error=getSensorMetadataByID_ms_stdev_q1]{data/query.dat};
\addplot+[ bar shift=-0.0cm,mark=none,fill=blue!60, draw=none,error bars/.cd,
      y dir=both,
      y explicit,error bar style={draw=blue}, error mark=|,error mark options={draw=blue}]table[x=Test,y=getSensorMetadataByID_ms_mean_q2,y error=getSensorMetadataByID_ms_stdev_q2]{data/query.dat};
\addplot+[  mark=none,bar shift=0.3 cm,fill=blue!60, draw=none,error bars/.cd,
      y dir=both,
      y explicit,error bar style={draw=blue}, error mark=|,error mark options={draw=blue}]table[x=Test,y=getSensorMetadataByID_ms_mean_q3,y error=getSensorMetadataByID_ms_stdev_q3]{data/query.dat};
\end{axis}
\end{tikzpicture}
\caption{Execution time for querying service.}
\label{fig:results_querying}
\end{figure}

As summarized in Figure \ref{fig:results_querying},  for both \emph{Filter$_1$} and \emph{Filter$_2$} queries, the service maintain a response latency always below 185 ms, while \emph{Aggregation} requires less than 85 ms. These values remains stable as data volume increases. This demonstrates an efficient query execution compared to typical federated SQL analytics, achieved through topic-based organization of records and optimized indexing strategies. 
Across all configurations, the Knowledge Graph query component (highlighted in blue in the Figure) introduces a consistent minimal overhead of approximately 10-15 ms, regardless of database size. 

We point out that in real-world scenarios, although long-term data storage is beneficial to enable historical data analysis and auditing, storage policies typically define a limited retention period for monitored data. 
Frequently adopted policies involves storing high-resolution, low-granularity data for a limited period (e.g., a few months), followed by storage of aggregated data (e.g., at hour level). For example, a 1 Hz sensor observations retained for six months, complemented by an additional 12 months of aggregated data would require 15.5M records, while a 100Hz sensor retained for 1 month, plus 12 months of aggregated data would require 29M records.
The largest databases used in the experimentation are therefore consistent  with this assumption.

\subsection{Stream processing latency}
\label{subsec:eval_platform}
The section is complemented with an assessment of the end-to-end latency of the platform, aimed to evaluate the processing time introduced by each component.
In this context, stream generators are assumed to produce messages that are then handled and processed by the following components: the preprocessing module, Kafka, a stream transformation pipeline, and finally republished to Kafka as a derived stream.
Experiments were conducted under these conditions:
\begin{itemize}
    \item input message rates of 100, 500, and 1000 messages per second, corresponding to low, moderate and high load conditions.
    \item All messages are assumed to be published in the same topic by multiple sensors.
    \item Each configuration was evaluated over multiple time windows, from 30 seconds to 10 minutes, i.e. \{30s, 60s, 120s, 300s, 600s\}, to verify performance stability over extended periods and detect potential degradation due to resource exhaustion or state accumulation. 
\end{itemize}
For each experiment, latency measurement was collected for each message at each stage of the platform pipeline. 
Average latency and standard deviations are computed 
over 10 runs.
First, a focus on the stream transformation pipeline will be presented, followed by the end-to-end analysis.
\paragraph{Stream transformation pipeline}
A first experiment focuses on latency for message processing within a stream transformation pipeline in Apache Flink. Hereby, we assume a pipeline consisting of a \emph{source} operator reading from a Kafka topic, a \emph{filter} operator that applies filtering conditions on specific message fields, a \emph{map} operator to remove unnecessary fields and finally a \emph{sink} operator to republish the derived stream  to Kafka on a new topic. 

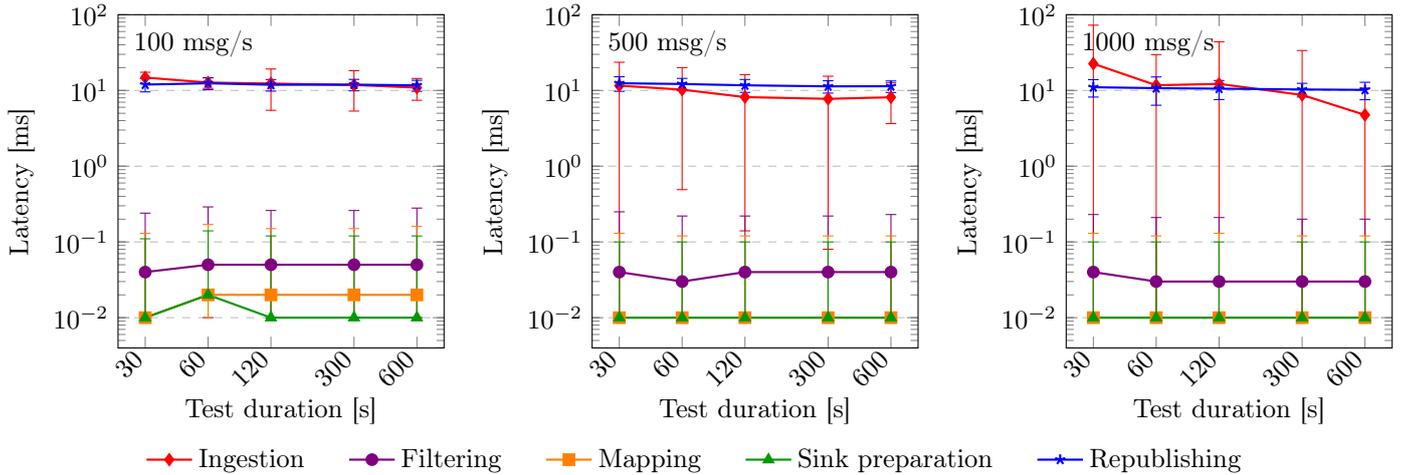
\begin{figure*}[t]
\centering

\begin{minipage}{0.32\textwidth}
\centering
\begin{tikzpicture}
\begin{axis}[
    width=\textwidth,
    height=6cm,
    xlabel={Test duration [s]},
    xlabel style={yshift=0.5em}, 
    ylabel={Latency [ms]},
    xticklabels={30, 60, 120, 300, 600},
    xticklabel style={rotate=45, anchor=east},
    xtick={1,2,3,4,5},
    xmode=log,
    log basis x=10,
    ymode=log,
    log basis y=10,
    xtick={30,60,120,300,600},
    ymin=0,
    ymax=100,
    legend style={draw=none},
    legend pos=north east,
    ymajorgrids=true,
    grid style=dashed,
    cycle list name=color list,
]
\addplot[color=red, mark=diamond*, thick, error bars/.cd, y dir=both, y explicit] 
table [x=X, y=Y, y error plus=Yplus, y error minus=Yminus] {
X     Y      Yplus   Yminus
30    14.81   2.59    2.59
60    12.71   2.05    2.05
120   12.34   6.87    6.87
300   11.80   6.46    6.46
600   10.90   3.48    3.48
};

\addplot[color=violet, mark=*, thick, error bars/.cd, y dir=both, y explicit] 
table [x=X, y=Y, y error plus=Yplus, y error minus=Yminus] {
X     Y      Yplus   Yminus
30     0.04   0.20    0.03
60     0.05   0.24    0.04
120    0.05   0.21    0.04
300    0.05   0.21    0.04
600    0.05   0.23    0.04
};

\addplot[color=orange, mark=square*, thick, error bars/.cd, y dir=both, y explicit] 
table [x=X, y=Y, y error plus=Yplus, y error minus=Yminus] {
X     Y      Yplus   Yminus
30     0.01   0.12    0.00
60     0.02   0.15    0.01
120    0.02   0.13    0.01
300    0.02   0.13    0.01
600    0.02   0.14    0.01
};

\addplot[color=green!60!black, mark=triangle*, thick, error bars/.cd, y dir=both, y explicit] 
table [x=X, y=Y, y error plus=Yplus, y error minus=Yminus] {
X     Y      Yplus   Yminus
30     0.01   0.10    0.00
60     0.02   0.12    0.00
120    0.01   0.11    0.00
300    0.01   0.11    0.00
600    0.01   0.11    0.00
};

\addplot[color=blue, mark=star, thick, error bars/.cd, y dir=both, y explicit] 
table [x=X, y=Y, y error plus=Yplus, y error minus=Yminus] {
X     Y      Yplus   Yminus
30    11.96   2.39    2.39
60    12.44   2.23    2.23
120   11.88   2.07    2.07
300   11.94   2.06    2.06
600   11.68   1.93    1.93
};
\node[anchor=north west] at (rel axis cs:0.02,0.98) {100 msg/s};
\end{axis}
\end{tikzpicture}
\end{minipage}
\hfill
\begin{minipage}{0.32\textwidth}
\centering
\begin{tikzpicture}
\begin{axis}[
    width=\textwidth,
    height=6cm,
    xlabel={Test duration [s]},
    xlabel style={yshift=0.5em}, 
    ylabel={Latency [ms]},
    xticklabels={30, 60, 120, 300, 600},
    xticklabel style={rotate=45, anchor=east},
    xtick={1,2,3,4,5},
    xmode=log,
    log basis x=10,
    ymode=log,
    log basis y=10,
    xtick={30,60,120,300,600},
    ymin=0,
    ymax=100,
    legend style={draw=none},
    legend pos=north east,
    ymajorgrids=true,
    grid style=dashed,
    cycle list name=color list,
]
\addplot[color=red, mark=diamond*, thick, error bars/.cd, y dir=both, y explicit] 
table [x=X, y=Y, y error plus=Yplus, y error minus=Yminus] {
X     Y      Yplus   Yminus
30    11.58  11.99   11.57
60    10.25   9.76    9.76
120    8.17   8.03    8.03
300    7.75   7.67    7.67
600    8.13   4.48    4.48
};

\addplot[color=violet, mark=*, thick, error bars/.cd, y dir=both, y explicit] 
table [x=X, y=Y, y error plus=Yplus, y error minus=Yminus] {
X     Y      Yplus   Yminus
30     0.04   0.21    0.03
60     0.03   0.19    0.02
120    0.04   0.18    0.03
300    0.04   0.18    0.03
600    0.04   0.19    0.03
};

\addplot[color=orange, mark=square*, thick, error bars/.cd, y dir=both, y explicit] 
table [x=X, y=Y, y error plus=Yplus, y error minus=Yminus] {
X     Y      Yplus   Yminus
30     0.01   0.12    0.00
60     0.01   0.11    0.00
120    0.01   0.11    0.00
300    0.01   0.11    0.00
600    0.01   0.11    0.00
};

\addplot[color=green!60!black, mark=triangle*, thick, error bars/.cd, y dir=both, y explicit] 
table [x=X, y=Y, y error plus=Yplus, y error minus=Yminus] {
X     Y      Yplus   Yminus
30     0.01   0.09    0.00
60     0.01   0.09    0.00
120    0.01   0.09    0.00
300    0.01   0.09    0.00
600    0.01   0.09    0.00
};

\addplot[color=blue, mark=star, thick, error bars/.cd, y dir=both, y explicit] 
table [x=X, y=Y, y error plus=Yplus, y error minus=Yminus] {
X     Y      Yplus   Yminus
30    12.45   2.71    2.71
60    12.17   2.26    2.26
120   11.69   2.26    2.26
300   11.33   2.10    2.10
600   11.37   1.97    1.97
};
\node[anchor=north west] at (rel axis cs:0.02,0.98) {500 msg/s};
\end{axis}
\end{tikzpicture}
\end{minipage}
\hfill
\begin{minipage}{0.32\textwidth}
\centering
\begin{tikzpicture}
\begin{axis}[
    width=\textwidth,
    height=6cm,
    xlabel={Test duration [s]},
    xlabel style={yshift=0.5em}, 
    ylabel={Latency [ms]},
    xticklabels={30, 60, 120, 300, 600},
    xticklabel style={rotate=45, anchor=east},
    xtick={1,2,3,4,5},
    xmode=log,
    log basis x=10,
    ymode=log,
    log basis y=10,
    xtick={30,60,120,300,600},
    ymin=0,
    ymax=100,
    legend style={draw=none},
    legend pos=north east,
    ymajorgrids=true,
    grid style=dashed,
    cycle list name=color list,
]

\addplot[color=red, mark=diamond*, thick, error bars/.cd, y dir=both, y explicit] 
table [x=X, y=Y, y error plus=Yplus, y error minus=Yminus] {
X     Y      Yplus   Yminus
30    22.54  50.24   22.53
60    11.71  18.05   11.70
120   12.22  31.70   12.21
300    8.72  24.94    8.71
600    4.75   6.03    4.74
};

\addplot[color=violet, mark=*, thick, error bars/.cd, y dir=both, y explicit] 
table [x=X, y=Y, y error plus=Yplus, y error minus=Yminus] {
X     Y      Yplus   Yminus
30    0.04   0.19    0.03 
60    0.03   0.18    0.02
120   0.03   0.18    0.02
300   0.03   0.17    0.02
600   0.03   0.17    0.02
};

\addplot[color=orange, mark=square*, thick, error bars/.cd, y dir=both, y explicit] 
table [x=X, y=Y, y error plus=Yplus, y error minus=Yminus] {
X     Y      Yplus   Yminus
30    0.01   0.12    0.01 
60    0.01   0.11    0.01
120   0.01   0.12    0.01
300   0.01   0.11    0.01
600   0.01   0.11    0.01
};

\addplot[color=green!60!black, mark=triangle*, thick, error bars/.cd, y dir=both, y explicit] 
table [x=X, y=Y, y error plus=Yplus, y error minus=Yminus] {
X     Y      Yplus   Yminus
30    0.01   0.09    0.01 
60    0.01   0.09    0.01
120   0.01   0.09    0.01
300   0.01   0.09    0.01
600   0.01   0.09    0.01
};

\addplot[color=blue, mark=star, thick, error bars/.cd, y dir=both, y explicit] 
table [x=X, y=Y, y error plus=Yplus, y error minus=Yminus] {
X     Y      Yplus   Yminus
30    11.03   2.85   2.85
60    10.73   4.35   4.35
120   10.56   2.98   2.98
300   10.31   2.11   2.11
600   10.18   2.64   2.64
};
\node[anchor=north west] at (rel axis cs:0.02,0.98) {1000 msg/s};
\end{axis}
\end{tikzpicture}
\end{minipage}

\vspace{0.3cm}
\begin{tikzpicture}
\begin{axis}[
    hide axis,
    xmin=0, xmax=1,
    ymin=0, ymax=1,
    legend style={
        draw=none,
        legend columns=5,
        /tikz/every even column/.append style={column sep=0.5cm}
    },
    legend to name=globallegend
]
\addlegendimage{color=red, mark=diamond*, thick}
\addlegendentry{Ingestion}
\addlegendimage{color=violet, mark=*, thick}
\addlegendentry{Filtering}
\addlegendimage{color=orange, mark=square*, thick}
\addlegendentry{Mapping}
\addlegendimage{color=green!60!black, mark=triangle*, thick}
\addlegendentry{Sink preparation}
\addlegendimage{color=blue, mark=star, thick}
\addlegendentry{Republishing}
\end{axis}
\end{tikzpicture}

\vspace{-0.3cm}
\centering
\ref{globallegend}

\caption{Stream transformation latency in Flink across different message rates and durations (log scale).}
\label{fig:Flink_performance}
\end{figure*}

Specific Flink configuration involved two critical parameters: 
the bundle size, which determines the maximum number of elements (e.g., events from a data stream) processed per batch, and the bundle time, that specifies the processing window duration (in milliseconds) before processing a bundle, setting an upper bound on processing delay.
Preliminary experiments determined the optimal trade-off between throughput and latency, so both parameters were configured to a value of 10 (bundle size = 10 elements, bundle time = 10 ms). This configuration balances efficient batching with minimal latency overhead, making it suitable for low-latency stream processing workloads.
Additional Flink configuration parameters included a buffer timeout of 50~ms, a parallelism of 4 task slots, 2GB of TaskManager memory, a network fraction of 0.1, and 4096 Network buffers.
Results, reported in Figure \ref{fig:Flink_performance}, demonstrate stable performance across all tested scenarios. The three experimental setups, i.e., 100 msg/s, 500 msg/s, and 1000 msg/s, are presented in the left, center, and right panels, respectively. The most significant variability occurs in the Input (orange) and Output (purple) stages, which can be attributed to Kafka consumer rebalancing. In contrast, the core computational operators, Filtering (blue) and Mapping (red), consistently execute in less than 0.5 ms. 

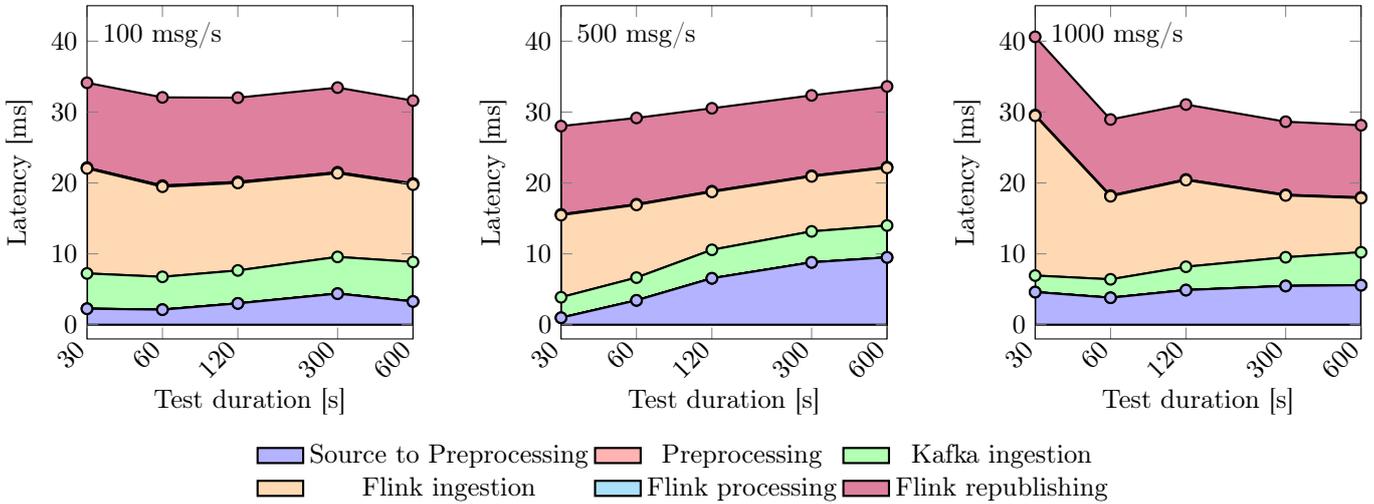
\begin{figure*}[t]
\centering

\begin{minipage}{0.32\textwidth}
\centering
\begin{tikzpicture}
\begin{axis}[
    width=\textwidth,
    height=6cm,
    stack plots=y,
    area style,
    enlarge x limits=false,
    xmin=30,
    xlabel={Test duration [s]},
    xlabel style={yshift=0.5em},  
    ylabel={Latency [ms]},
    xticklabels={30, 60, 120, 300, 600},
    xticklabel style={rotate=45, anchor=east},
    xmode=log,
    log basis x=10,
    xtick={30,60,120,300,600},
    ytick={0,10,20,30,40},
    yticklabels={0,10,20,30,40},
    ymin=-2,
    ymax=45,
    legend to name=total_latency_legend,  
    legend style={at={(0.5,-0.20)},anchor=north,draw=none},
    legend columns=3,
]
    \addplot [thick, mark=*, fill=blue!30] coordinates 
        {(30, 2.26) (60, 2.13) (120, 3.01) (300, 4.39) (600, 3.28)}
            \closedcycle;
    \addlegendentry{Source to Preprocessing}
    
    \addplot [thick, mark=*, fill=red!30] coordinates 
        {(30, 0.01) (60, 0.01) (120, 0.01) (300, 0.01) (600, 0.01)}
            \closedcycle;
    \addlegendentry{Preprocessing}
    
    \addplot [thick, mark=*, fill=green!30] coordinates 
        {(30, 4.97) (60, 4.62) (120, 4.64) (300, 5.17) (600, 5.57)}
            \closedcycle;
    \addlegendentry{Kafka ingestion}
    
    \addplot [thick, mark=*, fill=orange!30] coordinates 
        {(30, 14.81) (60, 12.71) (120, 12.34) (300, 11.80) (600, 10.90)}
            \closedcycle;
     \addlegendentry{Flink ingestion}
     
    \addplot [thick, mark=*, fill=cyan!30] coordinates 
        {(30, 0.12) (60, 0.16) (120, 0.15) (300, 0.14) (600, 0.17)}
            \closedcycle;
    \addlegendentry{Flink processing}
    
	\addplot [thick, mark=*, fill=purple!50] coordinates 
        {(30, 11.96) (60, 12.44) (120, 11.88) (300, 11.94) (600, 11.68)}
            \closedcycle;
    \addlegendentry{Flink republishing}
\node[anchor=north west] at (rel axis cs:0.02,0.98) {100 msg/s};    
\end{axis}
\end{tikzpicture}  
\end{minipage}
\hfill
\begin{minipage}{0.32\textwidth}
\centering
\begin{tikzpicture}
\begin{axis}[
    width=\textwidth,
    height=6cm,
    stack plots=y,
    area style,
    enlarge x limits=false,
    xmin=30,
    xlabel={Test duration [s]},
    xlabel style={yshift=0.5em},  
    ylabel={Latency [ms]},
    xticklabels={30, 60, 120, 300, 600},
    xticklabel style={rotate=45, anchor=east},
    xmode=log,
    log basis x=10,
    xtick={30,60,120,300,600},
    ytick={0,10,20,30,40},
    yticklabels={0,10,20,30,40},
    ymin=-2,
    ymax=45,
]
    \addplot [thick, mark=*, fill=blue!30] coordinates 
        {(30, 0.99) (60, 3.44) (120, 6.55) (300, 8.80) (600, 9.50)}
            \closedcycle;
    
    \addplot [thick, mark=*, fill=red!30] coordinates 
        {(30, 0.00) (60, 0.00) (120, 0.00) (300, 0.00) (600, 0.00)}
            \closedcycle;
    
    \addplot [thick, mark=*, fill=green!30] coordinates 
        {(30, 2.89) (60, 3.21) (120, 4.02) (300, 4.37) (600, 4.50)}
            \closedcycle;
    
    \addplot [thick, mark=*, fill=orange!30] coordinates 
        {(30, 11.58) (60, 10.25) (120, 8.17) (300, 7.75) (600, 8.13)}
            \closedcycle;
     
    \addplot [thick, mark=*, fill=cyan!30] coordinates 
        {(30, 0.11) (60, 0.10) (120, 0.10) (300, 0.10) (600, 0.11)}
            \closedcycle;
    
	\addplot [thick, mark=*, fill=purple!50] coordinates 
        {(30, 12.45) (60, 12.17) (120, 11.69) (300, 11.33) (600, 11.37)}
            \closedcycle;
\node[anchor=north west] at (rel axis cs:0.02,0.98) {500 msg/s};
\end{axis}
\end{tikzpicture} 
\end{minipage}
\hfill
\begin{minipage}{0.32\textwidth}
\centering
\begin{tikzpicture}
\begin{axis}[
    width=\textwidth,
    height=6cm,
    stack plots=y,
    area style,
    enlarge x limits=false,
    xmin=30,
    xlabel={Test duration [s]},
    xlabel style={yshift=0.5em},  
    ylabel={Latency [ms]},
    xticklabels={30, 60, 120, 300, 600},
    xticklabel style={rotate=45, anchor=east},
    xmode=log,
    log basis x=10,
    xtick={30,60,120,300,600},
    ytick={0,10,20,30,40},
    yticklabels={0,10,20,30,40},
    ymin=-2,
    ymax=45,
]
    \addplot [thick, mark=*,fill=blue!30 ] coordinates 
        {(30, 4.61) (60, 3.81) (120, 4.89) (300, 5.47) (600, 5.58)}
            \closedcycle;
    
    \addplot [thick,fill=red!30,mark=*] coordinates 
        {(30, 0.00) (60, 0.00) (120, 0.00) (300, 0.00) (600, 0.00)}
            \closedcycle;
    
    \addplot [thick,mark=*,fill=green!30] coordinates 
        {(30, 2.33) (60, 2.60) (120, 3.29) (300, 4.05) (600, 4.64)}
            \closedcycle;
    
    \addplot [thick,mark=*,fill=orange!30] coordinates 
        {(30, 22.54) (60, 11.71) (120, 12.22) (300, 8.72) (600, 7.65)}
            \closedcycle;
     
    \addplot [thick,mark=*, fill=cyan!30] coordinates 
        {(30, 0.10) (60, 0.10) (120, 0.10) (300, 0.09) (600, 0.09)}
            \closedcycle;
    
	\addplot [thick,mark=*, fill=purple!50] coordinates 
        {(30, 11.03) (60, 10.73) (120, 10.56) (300, 10.31) (600, 10.18)}
            \closedcycle;
\node[anchor=north west] at (rel axis cs:0.02,0.98) {1000 msg/s};
\end{axis}
\end{tikzpicture} 
\end{minipage}

\vspace{0.1cm}
\centering
\ref{total_latency_legend}

\caption{End-to-end latency measurements across different message rates and durations.}
\label{fig:end_to_end_perfomance}
\end{figure*}

\paragraph{End-to-end latency}
A second experiment aims to evaluate the end-to-end latency of the platform.
Figure \ref{fig:end_to_end_perfomance} presents results as a stacked area line chart,  illustrating the contribution of each pipeline stage to total latency. The cumulative profiles reveal that the latency between message generation and publication to Kafka is on the order of 10 ms, and represents the baseline overhead introduced by the data ingestion layer. Such latency is negligible for most industrial monitoring applications.

Input/Output stages dominate the overall latency, including message ingestion, (de)serialization, buffering, and network transfer operations. The measured latency ranges from 3–10 ms for the Source to Pre\-processing module, 8–11 ms for the Kafka - Flink stage (Flink ingestion), and 10–12 ms for the final Flink republishing to Kafka.

 Finally, the processing (CPU-bound) stages, including  the preprocessing and the Flink pipeline, introduce minimal overhead, as already demonstrated.

 Overall, the input rate did not produce a proportional increase in latency, showing that the system scales horizontally and maintains consistent performance under moderate load growth.

\section{Conclusion}
\label{sec:conclusion}
This paper presented a context-aware semantic approach for data stream management in Industrial IoT environments, combining Knowledge Graphs, semantic reasoning, and real-time stream processing technologies. 
The approach formalizes the core abstractions of stream management, within an ontology-driven model that enables interoperability, reasoning, and dynamic access control across heterogeneous industrial data. By embedding de\-cla\-ra\-tive semantics into the data processing flow, the platform bridges the gap between low-level stream manipulation and high-level contextual awareness (research questions 1-2).
The platform integrates stream processing (through Kafka and Flink) and Knowledge Graph reasoning, supporting flexible, secure, and semantically interoperable data workflows, with role- and context-aware access control that adapts to users’ operational context.
Experimental results demonstrated high scalability and low-latency performance, confirming the feasibility of real-time semantic reasoning in distributed stream environments, even with commodity hardware (research question 3).

Future work will address several research directions. 
First, we aim at supporting formal validation and type-safe composition of complex stream pipelines. Additionally, we aim to investigate the automated discovery of semantic relationships to further reduce manual configuration in increasingly large-scale and dynamic IoT environments.
Finally, we aim to integrate Machine Learning models for adaptive stream optimization, leveraging semantic metadata as structured contextual knowledge for model configuration and retraining. In this direction, a recent result is the use of Large Language Models (LLMs) and Retrieval-Augmented Generation (RAG) approaches to enhance information retrieval and decision support for employees based on natural language requests, which are then translated to graph queries to extract contextually-related information \cite{10.1145/3762669}. 







\section*{Declaration of competing interest}
The authors declare that they have no known competing financial interests or personal relationships that could have appeared to influence the work reported in this paper.

\section*{Data availability}
Code and graph schemas are available at the project's GitHub repository https://github.com/Homey-Prin22/framework.

\section*{Acknowledgments}
\begin{figure}[H]
\includegraphics[width=3cm]{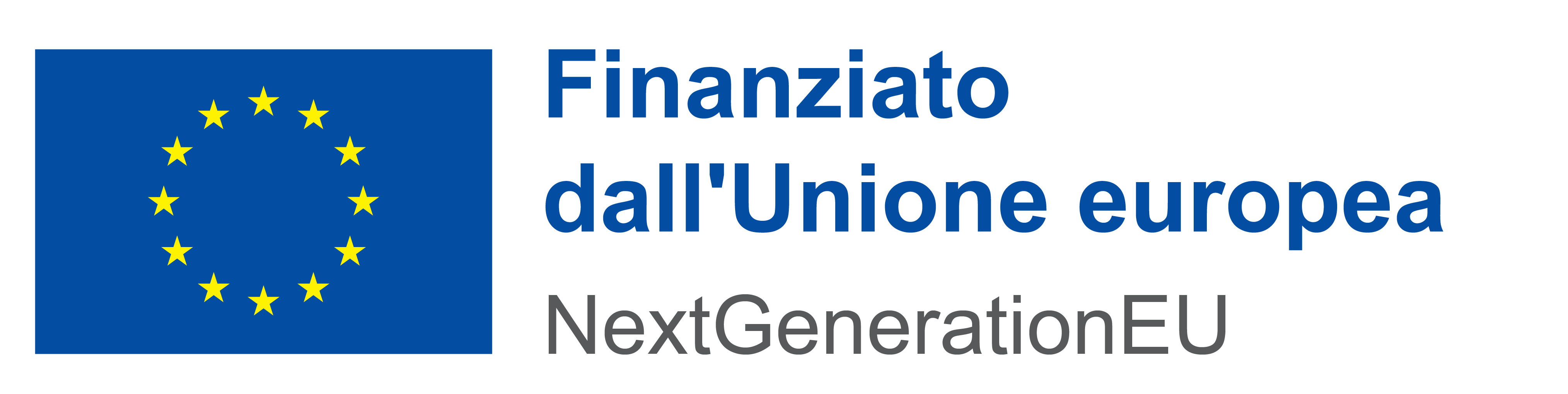}%
\end{figure}
\noindent
This work has been partially supported by the PRIN 2022 project ``HOMEY: a Human-centric IoE-based Framework for Supporting the Transition Towards Industry 5.0'' (code: 2022NX7WKE, CUP: F53D23004340006), funded by the European Union - Next Generation EU, Mission 4 Component 1.

\bibliographystyle{elsarticle-num}
\bibliography{bibliography}

\end{document}